# Non-linear Optical Spectroscopy of Excited Exciton States for Efficient Valley Coherence Generation in $WSe_2$ Monolayers

G. Wang, X. Marie, I. Gerber, T. Amand, D. Lagarde, L. Bouet, M. Vidal, A. Balocchi & B. Urbaszek

*Université de Toulouse, INSA-CNRS-UPS, LPCNO,*

*135 Av. de Rangueil, 31077 Toulouse, France*

**Monolayers (MLs) of $MoS_2$ and $WSe_2$ are 2D semiconductors with strong, direct optical transitions that are governed by tightly Coulomb bound electron – hole pairs (excitons). The optoelectronic properties of these transition metal dichalcogenides are directly related to the inherent crystal inversion symmetry breaking. It allows for efficient second harmonic generation (SHG) and is at the origin of chiral optical selections rules, which enable efficient optical initialization of electrons in specific K-valleys in momentum space. Here we demonstrate how these unique non-linear and linear optical properties can be combined to efficiently prepare exciton valley coherence and polarization through resonant pumping of an excited exciton state. In particular a new approach to coherent alignment of excitons following two-photon excitation is demonstrated. We observe a clear deviation of the excited exciton spectrum from the standard Rydberg series via resonances in SHG spectroscopy and two- and one-photon absorption. The clear identification of the 2s and 2p exciton excited states combined with first principle calculations including strong anti-screening effects allows us to determine an exciton binding energy of the order of 600 meV in ML $WSe_2$.**

Transition metal dichalcogenide (TMDC) monolayers (MLs) such as $WSe_2$ and $MoS_2$ are an exciting class of atomically flat, two-dimensional materials for electronics and optoelectronics[1-7]. Inversion symmetry breaking together with the spin-orbit interaction lead to a unique coupling of carrier spin and k-space valley physics for new device functionalities. The circular polarization ($\sigma^+$ or $\sigma^-$ ) of the absorbed or emitted photon can be directly associated with selective carrier excitation in one of the two non-equivalent K valleys (K+ or K-, respectively) [8-12], see figure 1b,c. The chiral optical selection rules open up very exciting possibilities of manipulating carriers in valleys with contrasting Berry phase curvatures[13]. Initially these unique spin and k-valley phenomena have been described in a single particle picture. In reality electrons and holes will be bound by the strong Coulomb interaction and form excitons which in a standard 2D system like GaAs quantum wells can be described by a Rydberg series, analogous to the hydrogenic system, with even ns and odd np states (n=1, 2, 3…). Large exciton binding energies of few hundreds of meV are expected due to the rather large effective electron mass and small effective dielectric constant[14-17]. The potentially strong impact of exciton states on the new physics related to valley index manipulation and non-linear optical effects has not been revealed yet and the exciton binding energy has not been measured so far for ML $WSe_2$.

Here we combine the energy and the high polarization selectivity of non-linear and linear optical spectroscopy to first uncover the excited exciton spectrum which does not follow the usual Rydberg series and second, excite the exciton resonances for controlled valley state preparation. We clearly distinguish between absorption maxima associated with effects of exciton-phonon coupling and excited exciton states (2s, 2p,…)[18-20]. The position of the exciton states and the free carrier bandgap are affirmed by one-photon and two-photon optical spectroscopy together with *ab initio GW*-Bethe Salpeter Equation calculations[21] and allow an extrapolation of the exciton binding energy. We observe additional excitonic resonances by performing second harmonic generation (SHG) excitation spectroscopy, which allows to tune the frequency doubling efficiency over three orders of magnitude. Following resonant excitation of the excited exciton states we demonstrate strong exciton valley polarization. Our measurements show that significant exciton valley coherence[22], corresponding to the generation of a coherent superposition of two exciton spin states from the K+ and K- valleys, can be created with a *two-photon* process. This demonstrates that the exciton alignment can be controlled with an excitation energy much lower than the band gap. None of these effects were demonstrated before in 2D or quasi-2D systems.

We studied $WSe_2$ flakes obtained by micro-mechanical cleavage of bulk $WSe_2$ crystal (from 2D Semiconductors, USA) on 90 nm $SiO_2$ on a Si substrate. The 1ML region is identified by optical contrast (figure 1a) and by photoluminescence (PL) spectroscopy. We have measured the two photon excitation spectra using picosecond pulses generated by a tuneable optical parametric oscillator (OPO) synchronously pumped by a mode-locked Ti:Sa laser. For the one-photon PL excitation experiments the excitation is provided by the frequency doubled OPO pulses (see Methods). In the experiments presented here the light is propagating perpendicular to the 2D layer plane. Thus the selection rules for interband transitions impose that the one-photon absorption occurs on the ns exciton states whereas np states will be allowed for two photon transitions (n is an integer)[18,19].

Figure 1e displays the results of two-photon optical spectroscopy experiments performed on the $WSe_2$ ML for a linearly polarized excitation laser with energy $E_{laser}$ = 0.946 eV, much lower than the optical gap of around ~1.7eV. The narrow line at E = 1.893 eV= 2x$E_{laser}$ corresponds to the laser Second Harmonic Generation (SHG) in the $WSe_2$ ML. The PL components at lower energy at E= 1.75 eV and E= 1.72 eV are the neutral and charged exciton (trion) spectra respectively. Under linearly polarized laser excitation, only the highest energy peak shows linear polarization in emission and is therefore ascribed to the neutral exciton ground state $X_A^{1s}$ (simple bound electron-hole pair), as a coherent superposition of valley exciton spin states is created[22,23]. As expected both the $X_A^{1s}$ luminescence intensity following two-photon absorption and the SHG intensity increase quadratically with the laser excitation power (see inset in figure 1e)[24-26].

Here we focus on the optical spectroscopy of the neutral $X_A$ exciton in ML $WSe_2$ which is characterized by a luminescence spectrum with FWHM~10 meV at T=4 K. This is much narrower than the spectra measured so far for $MoS_2$ MLs [4,11]. This allows a precise determination of the exciton excited state spectra both for one or two photons optical spectroscopy experiments. An additional advantage is that the maximum of the valence band at the Γ point lies more than 500 meV below the K point[27] whereas for ML $MoS_2$ this energy difference is just a few tens of meV according to recent calculations [14,28]. When performing the excitation spectra in ML $WSe_2$, the excitation of indirect transitions should thus not energetically overlap with key features associated to the direct transition of excited exciton states.

**Two-photon absorption experiments**. The dependence of the neutral exciton $X_A^{1s}$ luminescence intensity as a function of the laser excitation energy is plotted in figure 2a,c. In this two-photon Photoluminescence Excitation (2P-PLE) experiment, the laser power is kept constant and the detection energy is set to the emission peak of $X_A^{1s}$. The intensity variation of the 2P-PLE curve displays two main features: a clear peak at ~1.898 eV and a clear monotonous increase in the high energy region 2.35 eV. The first energy peak labeled $X_A^{2p}$, 140 meV above the $X_A^{1s}$ exciton emission, corresponds to the 2p exciton state absorption allowed for two-photon optical spectroscopy (the $X_A^{1s}$ absorption is forbidden by the two photons section rules).

The vertical arrow labeled $E_{gap}$=2.4 eV corresponds to the calculated energy of the onset of the continuum states (free electron-hole absorption) as evidenced by the quasi-particle gap in figure 5.a, calculated at the $GW_0$ level including spin-orbit coupling, see Methods. This value is slightly smaller than a recent theoretical estimate[29] (2.50 eV), the discrepancy can be attributed to differences in computational settings. If we define the exciton binding energy as $E_{gap}$-E($X_A^{1s}$)=$E_b$ we can estimate $E_b$ ~ 600±50 meV. A high neutral exciton binding energy is consistent with the measured charged exciton binding energy in figure 1d,e : $E_b$ (trion)=35 meV[38,39]. The calculated value $E_{gap}$=2.4 eV explains very well the 2P-PLE signal increase in the 2.3-2.35 eV region. This increase cannot be assigned to the absorption of an exciton state associated to a Van Hove singularity[17] since the absorption on its ground state is forbidden in a two-photon process.

Remarkably a detailed analysis of the spectra around the $X_A^{2p}$ absorption region shows two secondary peaks located ~30 and 60 meV above the 2p main exciton peak (see figure 2a and SI). We interpret these secondary peaks as the result of the enhancement in the ground-state PL signal for excitation energies permitting resonant phonon relaxation down to the $X_A^{2p}$ states. Indeed Raman studies identified two optical phonon modes with energy of about $E_{phonon}$ =31 meV (~250 $cm^{-1}$) in $WSe_2$ ML (see SI)[30,31] . The mode labeled $E^1_{2g}$ is strong in one-photon absorption, whereas the mode labelled $A_{1g}$ is more likely to play a role in two-photon processes for symmetry reasons (see SI). Because the resonances are much broader than the spectral width of the exciting laser (see SI), the peaks observed are not attributed to simple inelastic scattering of the laser line (i.e., Raman lines), but rather to enhanced PL when the laser energy corresponds to $X_A^{2p}$ + $E_{phonon}$ and $X_A^{2p}$ + 2. $E_{phonon}$ for the peaks observed in figure 2a at ~1.93 and ~1.96 eV respectively. We emphasize that we never observed any Raman peak with a possible phonon energy of ~140 meV, in agreement with the experimental and theoretical Raman studies. This confirms that the peak labeled $X_A^{2p}$ is a real excited exciton state and not a phonon resonance.

**Second Harmonic Generation spectroscopy.** To uncover additional exciton resonances we employ a different non-linear optical technique with distinctly different polarization selection rules due to the intricate interplay between the electric and magnetic dipole interaction (see SI). We have performed SHG spectroscopy of excitons, a very powerful yet unexplored technique for 2D semiconductors[25,32,33], based on the drastic enhancement of the SHG efficiency when in resonance with an excitonic transition. Figure 3 displays the SHG intensity as a function of the SHG energy which probes directly the non-linear two-photon dielectric susceptibility: $d_{11}(2\omega; \omega,\omega)$, see SI. A clear peak is observed at about 1.9 eV, similarly to the 2P-PLE curve in figure 2, confirming the spectral position of the $X_A^{2p}$ exciton state[34]. A clear resonant signal is also evidenced at E=2.03 eV. We tentatively assign this peak to the $X_A^{3p/3s}$ state though the simulated spectrum does not present a clear feature in this region. The theoretical determination of these high energy exciton excited states, using beyond mean-field approaches is very challenging since it raises serious problem of convergence[17,35], due to their highly delocalized character.

In contrast to the 2P-PLE experiment where the $X_A^{1s}$ and $X_B^{1s}$ absorption are forbidden (and indeed absent in figure 2), we observe clearly in the SHG curve their strong resonances at E= 1.75 eV and E= 2.17 eV, respectively. From this measurement we get the difference between the SO energy splitting in the conduction and the valence band of ~420 meV in agreement with the one photon-PLE results (figure 4c), the reflectivity spectra[23] and our calculated value of the SO splitting, see figure 5a, in good agreement with a previous study[16]. The SHG signal at the $X_A^{1s}$ resonance is three orders of magnitude stronger than away from an exciton resonance, an important finding for applications of ML WSe2 in non-linear optics. Interestingly we observe in figure 3b a spectral component ~140 meV above the $X_B^{1s}$ resonance which could be related to the $X_B^{2p}$ state. This means the energy splitting between the first excited exciton state and its ground state is similar for the $X_A$ and $X_B$ exciton. This result is indeed expected as the A and B valence states have very similar effective masses (see figure 5a).

**One-photon absorption spectroscopy.** The findings from the two polarization resolved non-linear optical techniques are further complemented by one photon optical spectroscopy. In this case the polarization selection rules give access to the s-exciton states whereas the p-states are forbidden. Figure 4 displays the one photon PL Excitation (1P-PLE) results. In this case the luminescence intensity of the 1s neutral exciton is a linear, and not quadratic, function of the laser intensity. A clear peak corresponding to the $X_A^{2s}$optically allowed absorption is observed at ~1.9 eV, 140 meV above the ground state $X_A^{1s}$ .

Comparing the results of 1P-PLE and 2P-PLE in figure 2 and 4 we can infer that the $X_A^{2s}$ and $X_A^{2p}$exciton states have the same energy within the spectral resolution of our excitation spectra (which is of the order of 5 meV )[36,37]. An energy splitting $X_A^{2p} - X_A^{2s}$ is possible because of the combined effects of the spin-orbit and exciton exchange interactions. We also observe in figure 4a an additional peak at 1.93 eV, about 30 meV above the $X_A^{2s}$ absorption energy, which could be assigned to the phonon resonances similarly to the one identified in the 2P-PLE (figure 2a).

**Efficient generation of exciton valley coherence in two-photon absorption.** A key argument for the attribution of the observed maxima in one and two-photon absorption to excitonic states comes from the simultaneous increase of the PL polarization of the 1s exciton states. Following excitation with linearly polarized light, the $X_A^{1s}$exciton emission is linearly polarized but its amplitude globally decreases when the laser energy increases (figure 2a,c and 4a,c), except when in resonance with an excited exciton state. We emphasize that this observation of exciton alignment is independent of the direction of the incident laser polarization (see figure 2d), which confirms that the observed linear polarization is not due to an in-plane asymmetry inducing exciton anisotropic exchange interaction as for many 1D and 0D semiconductor systems[44]. In contrast to the exciton spin coherence evidenced in resonant excitation of GaAs quantum wells[45,46], here the linear polarization probes a spin coherence arising from excitons from two different valleys[22] . The results presented in figure 2 demonstrate for the first time in a semiconductor that significant exciton spin coherence can be created with a two-photons process. It is clear from figures 2a,c and 4a,c that the different resonances in intensity associated to the 2p and 2s exciton absorption are accompanied by a clear enhancement of the exciton PL linear polarization which is multiplied by a factor of more than two. Therefore the photogeneration of a coherent superposition of valley K+ and K- exciton states is enhanced when the absorption occurs on one of its excited exciton states. Typically the valley coherence is stronger when the excited state $X_A^{2s}$ (or $X_A^{2p}$ ) is directly photogenerated compared to the situation of the photogeneration of a hot $X_A^{1s}$ exciton state at the same energy. This probably results from a faster intra-exciton energy relaxation (2s→1s or 2p→ 1s) preserving the initially created coherent superposition of states. Note that the experiments on optical generation of excitonic valley coherence in ML $WSe_2$ recently published use a one-photon laser energy which coincides exactly with the 2s exciton absorption identified in our work[22]. In contrast to the coherent exciton alignment with the excitation laser polarization, the second harmonic generation polar plot in figure 2d and 2e is simply given by the underlying lattice symmetry[25,26].

Remarkably also the valley polarization is strongly enhanced when the circularly polarized excitation energy is resonant with the excited exciton state $X_A^{2s}$ or $X_A^{2p}$, as probed by the PL circular polarization in the 1P-PLE and 2P-PLE experiments shown in figure 2a and 4a, respectively. A clear polarization peak is observed for an energy 140 meV above the $X_A^{1s}$ exciton ground state energy. So the valley index initialization, which has been shown to be less and less efficient when the photogeneration energy increases[9-11,47] can be recovered when the excited exciton states 2s or 2p are directly photogenerated with one or two-photon excitation process, respectively.

**Exciton binding energy from experiment and theory.** The measured energy difference between $X_A^{1s}$ (1.75) and $X_A^{2s}$ (1.90 eV) is quite close to the calculated one of 0.25 eV as shown in figure 5b based on *GW*-Bethe-Salpeter equation, see Methods. Note that the calculation of the absorption spectrum in figure 5 was performed on a fine (21x21x1) **k**-point grid, an essential requirement to highlight the excited exciton states $X_A^{2s}$ which were missed in most of the previous calculations[15,17,35].

For the two-dimensional Wannier-Mott exciton the binding energy of the ground state 1s exciton writes simply[36] $E_b = \frac{(2\mu e^4)}{(\varepsilon\hbar)^2}$, where μ is the exciton reduced mass, ε the effective dielectric constant and e the electron charge. Using $1/\mu=1/m_e+1/m_h$, with $m_e=0.32.m_0$ and $m_h=0.35.m_0$ ($m_0$ is the free electron mass) as deduced from the calculated dispersion curves in the K valley[16] (figure 5a), we get an order magnitude estimation for the binding energy of $E_b$ =600 meV with an effective dielectric constant $\varepsilon = 5$. If we assume a simple hydrogenic Rydberg model the energy separation between n=2 and n=1 exciton s-state would be equal to $8.E_b/9$~440 meV. It is in strong contradiction with the experimental data presented above which shows that this energy splitting (~140 meV) is about 3 times smaller, and also in contradiction with $GW_0$-BSE estimate of this energy difference. Both our experiments and theory demonstrate clearly that the simple hydrogenic model fails to predict the energy of the different exciton states in the TMDC MLs as predicted in ref [17] and very recently observed in $WS_2$ MLs[40,41]. In the well known quasi 2D GaAs quantum wells the exciton state series can be well reproduced by the hydrogenic Rydberg model since the 2D layer is surrounded by a barrier material with very similar dielectric constant[18,19]. In contrast the investigated 2D crystals based on TMDC have an environment perpendicular to the layer with a much smaller dielectric constant than the 2D material (vacuum on one side and $SiO_2$/Si on the other side). For the exciton ground state, the average distance between the electron and the hole is so short that the exciton wavefunction is mainly located in the $WSe_2$ layer and the binding energy can be obtained with a simple formula for $E_b$ with an effective dielectric constant close to the bulk $WSe_2$ value[42]. The excited exciton states such as $X_A^{2s}$ or $X_A^{2p}$ are characterized by a longer average distance between the electron and the hole. As a consequence the carrier wavefunctions delocalize beyond the $WSe_2$ 2D layer and experience a weaker dielectric constant i.e. stronger Coulomb interaction. The effects of screening/anti-screening associated to the spatial variation of the dielectric constant[43] are in principle well taken into account in *ab initio* calculations, assuming a large separation between periodic images. The strong dependence of $E_b$ on (anti-)screening explains the remaining difference between the experimental results and the calculations for $X_A^{1s}$ and $X_A^{2s}$ in our work. Also, the calculation neglects the influence of the $SiO_2$/Si substrate used in the experiment. Despite the remaining discrepancies, the calculations confirm the strong deviation from the standard hydrogen Rydberg series observed here in the experiment for ML $WSe_2$.

In conclusion we have measured the spectral dependences of both the first order dielectric susceptibility (one-photon PLE) and the second order dielectric susceptibility (SHG excitation spectra for its real part and two-photon PLE for its imaginary part). These experiments directly demonstrate that excitonic effects enhance the linear and non-linear optical response of ML $WSe_2$ by several orders of magnitude. These findings can also find application in $MoS_2$, $WS_2$, $MoSe_2$ monolayers and TMDC heterostructures[1]. Our results not only reveal that the valley coherence can be achieved following two-photons excitation but also yield the determination of the exciton binding energy which is a crucial parameter for the optoelectronics and valley application of transition metal dichalcogenide 2D structures.

METHODS :

**Experiments** : The measurements are performed at T=4 K in a confocal microscope optimized for polarized PL experiments[11,44]. The detection spot diameter is ≈ 1 μm. The time-integrated PL emission is dispersed in a spectrometer and detected with a Si-CCD camera.

For the 2P-PLE, the flake is excited by picosecond pulses generated by a tunable optical parametric oscillator (OPO) synchronously pumped by a mode-locked Ti:Sa laser. The wavelength can be tuned between 1 and 1.6 μm. The typical pulse and spectral width are 1.6 ps and 3 meV respectively; the repetition rate is 80 MHz. For the 1P-PLE the excitation is provided by the frequency doubled OPO pulses. In that case the laser power has been kept in the μW range in the linear absorption regime. The PL circular polarization $P_c$ is defined as $P_c = (I_{\sigma+} - I_{\sigma-})/(I_{\sigma+} + I_{\sigma-})$. Here $I_{\sigma+}$ ( $I_{\sigma-}$) denotes the intensity of the right (σ+) and left (σ−) circularly polarized emission, analyzed by a quarter-wave plate placed in front of a linear polarizer. The PL linear polarization $P_L$ is defined as $P_L = (I_{//} - I_{\perp})/(I_{//} + I_{\perp})$. $I_{//}$ ($I_{\perp}$) denotes the intensity of the linearly polarized emission co- and cross-polarized with the laser.

**Computational details :**

Our calculations are performed using the Vienna *ab initio* simulation package[49]. The electron wave function is expanded in a plane wave basis set with an energy cutoff of 400 eV. Investigations of the quasiparticle band structure are done at the $GW_0$ level[50], with 2 updates of the $G$ term, on a (21x21x1) Γ-centered-grid **k**-point mesh. Optical absorption spectra are calculated on the BSE-$GW_0$ level, using the Tamm-Dancoff approximation, after careful studies of convergence issues with respect to Brillouin zone sampling, vacuum heights and number of included virtual states, see SI. Vacuum heights of at least 12 Å between periodic images of the monolayer and a lattice parameter of 3.28 Å (experimental values) have been used to define the calculation cell. Using these settings, we estimate that the precision of our calculated exciton binding energy is around 0.1 eV, due to uncertainty on the band-gap estimate and relative position of peaks in the simulated spectrum. Note that (i) the SO interaction has not been considered in the calculation of figure 5b and (ii) no conclusion can be drawn for exciton states above $X_A^{2s/2p}$ since finer reciprocal space meshes are mandatory but too computationally demanding to estimate the kernel.

*Acknowledgments :* We thank M. M. Glazov and B.L. Liu for stimulating discussions. We acknowledge partial funding from ERC Starting Grant No. 306719 and Programme Investissements d'Avenir ANR-11-IDEX-0002-02, reference ANR-10-LABX-0037- NEXT. I. Gerber thanks the CALcul en Midi-Pyrénées initiative (CALMIP, Grant 2013/14-P0812) for generous allocations of computational time.

**Figure Captions:**

**Figure 1**

a. Monolayer (ML) $WSe_2$ structure with broken inversion symmetry. The ML is identified by optical contrast measurements as is marked by the dashed rectangle.

b. Optical valley selection rules for circularly polarized laser excitation in the single particle picture for ML $WSe_2$.

c. Optical valley selection rules in the exciton representation. For simplicity, only the A-exciton series is shown.

d. Optical valley coherence generation following absorption of one, linearly polarized photon. The neutral exciton $X_A^{1s}$and the trion (T) transtions are marked. PL spectra co-polarized (cross) with the laser are shown in blue (green). The onset of the localised state emission can be seen at low energy.

e. Optical valley coherence generation via two-photon absorption at T=4K. The neutral exciton $X_A^{1s}$ and the trion (T) transtions are marked. PL spectra co-polarized (cross) with the laser are shown in blue (green).

f. Resonances in two-photon absorption intensity at T=4K corresponding to the $X_A^{2p}$ transition and phonon assisted absorption at energies $X_A^{2p}$ + $E_{phonon}$ and $X_A^{2p}$ + 2. $E_{phonon}$.

**Figure 2**

**Two-photon absorption spectroscopy of excited excitons for valley coherence initialization**.

a. Top panel. The intensity of the neutral exciton PL is plotted as a function of two times the excitation laser energy. The local maximum 140meV above the $X_A^{1s}$ emission is identified as $X_A^{2p}$. Phonon enhanced absorption is marked by dashed lines. Middle panel. The linear polarization of the $X_A^{1s}$ emission indicating exciton valley coherence is strongly enhanced when the two-photon excitation is in resonance with $X_A^{2p}$. Lower panel. The circular polarization of the $X_A^{1s}$ emission due to exciton valley polarization is strongly enhanced when the two-photon excitation is in resonance with $X_A^{2p}$.

b. Schematics of the two-photon absorption process, the subsequent relaxation and PL emission.

c. Same as figure a (top and middle panel) but over a wider laser energy range (black squares – PL intensity, blue circles – PL linear polarization). The $X_A^{2p}$ and the possible onset of free carrier absorption is indicated. The vertical arrow indicates the calculated gap energy $E_{gap}$=2.4 eV (see text).

d. Intensity of the $X_A^{1s}$ emission following two-photon absorption at 1.893 eV (red circles). The $X_A^{1s}$ polarization plane follows the excitation laser (green arrow) and is not linked to a specific lattice axis or symmetry, which confirms the generation of neutral exciton valley coherence. The second harmonic generation (black squares) polar plot is given by the underlying lattice symmetry.

e. Second harmonic generation as a function of the angle of the laser polarization with sample angle (black squares). Neutral exciton polarization (red circles) follows laser polarization.

**Figure 3**

**Second harmonic generation spectroscopy of exciton states.**

a. Schematics of SHG signal generation for resonant two-photon excitation of the $X_A^{2p}$.

b. Intensity of SHG as a function of two-photon laser excitation, the clear exciton resonances are marked.

**Figure 4**

**One-photon photoluminescence excitation spectroscopy.**

a. Top panel. The intensity of the neutral exciton PL is plotted as a function of the excitation laser energy. The local maximum 140meV above the $X_A^{1s}$ emission is identified as $X_A^{1s}$. Phonon enhanced absorption is marked by dashed lines. Middle panel. The linear polarization of the $X_A^{1s}$ emission due to exciton valley coherence is strongly enhanced when the one-photon excitation is in resonance with $X_A^{2s}$. Lower panel. The circular polarization of the $X_A^{1s}$ emission due to exciton valley polarization is strongly enhanced when the one-photon excitation is in resonance with $X_A^{2s}$.

b. Schematics of the one-photon absorption process and the subsequent relaxation and PL emission.

c. Same as figure a (top and middle panel) but over wider laser energy range (black squares – PL intensity, blue circles – PL linear polarization). The $X_A^{2s}$ and the $X_B^{1s}$ resonances are indicated.

**Figure 5**

a. Quasiparticle band structure of $WSe_2$ monolayer at the $GW_0$ level, including spin-orbit coupling perturbatively.

b. Frequency dependent imaginary part of the dielectric function $\varepsilon_2$ including excitonic effects and oscillator strengths of the optical transitions (bars), calculated at the BSE-$GW_0$ level without spin-orbit coupling. The corresponding $GW_0$ gap (2.4 eV) is also indicated.

## Figure 1

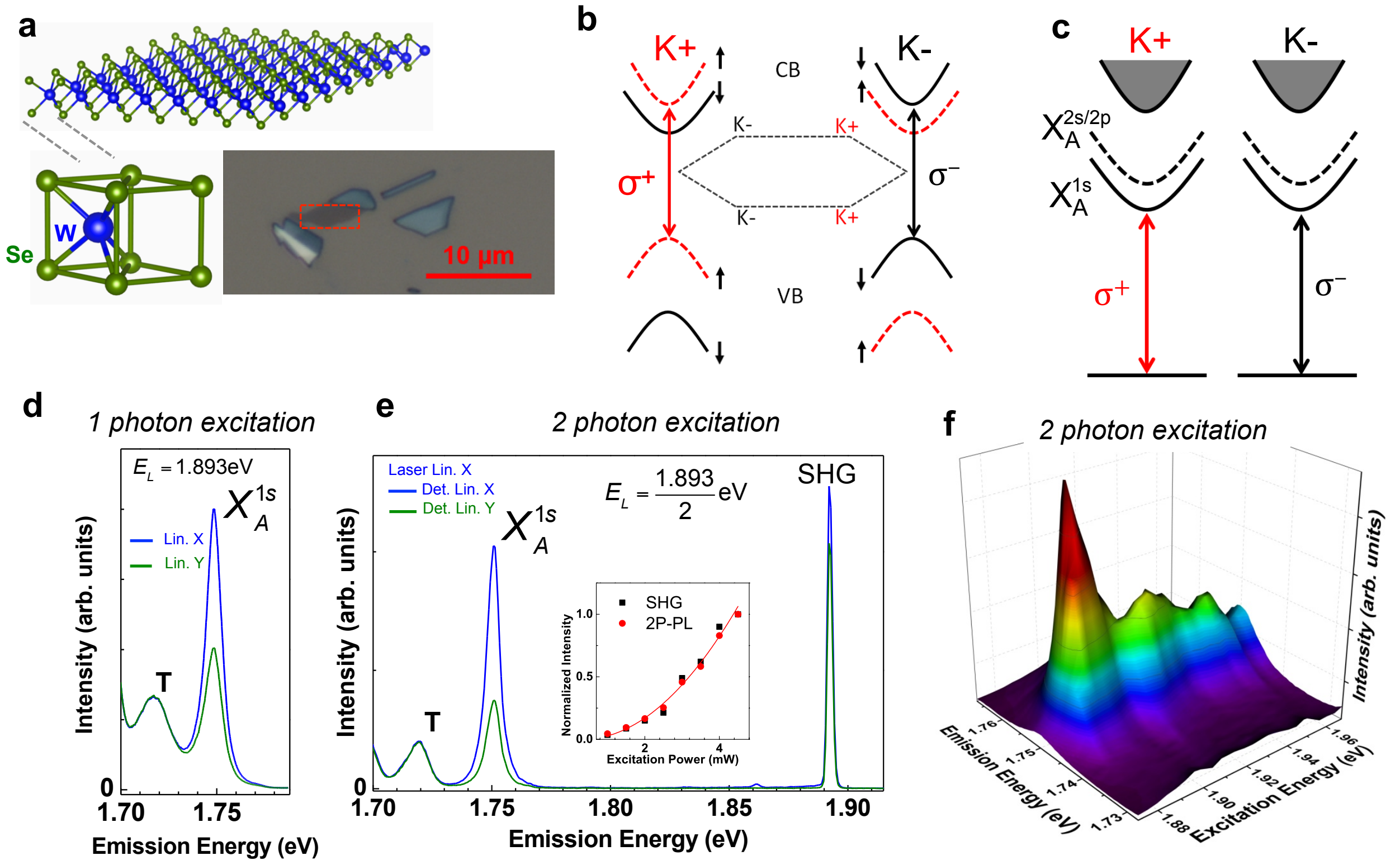

**Figure 2**

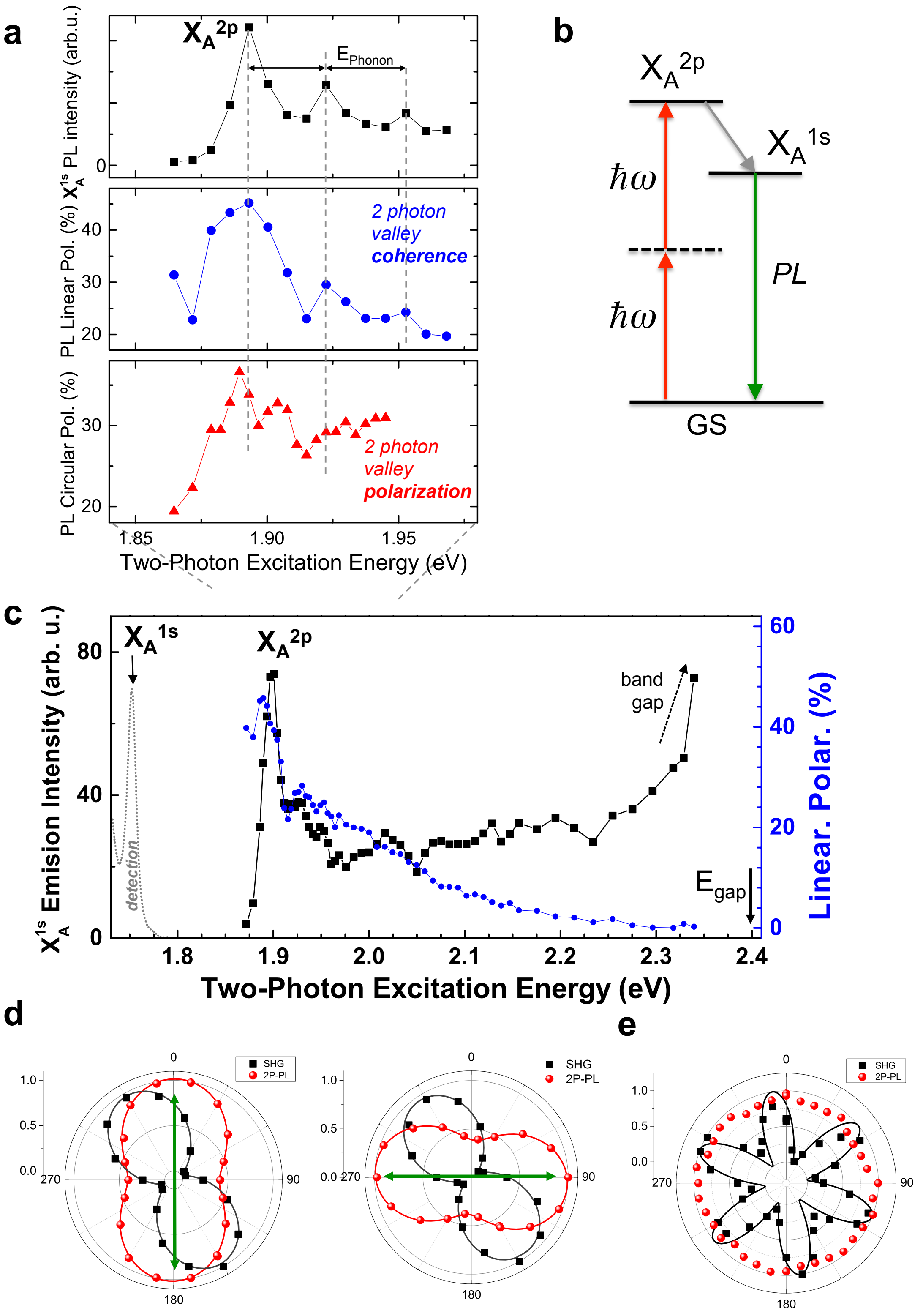

a
$X_A^{2p}$
$E_{Phonon}$
$X_A^{1s}$ PL intensity (arb.u.)
PL Linear Pol. (%)
2 photon valley coherence
PL Circular Pol. (%)
2 photon valley polarization
Two-Photon Excitation Energy (eV)
b
$X_A^{2p}$
$X_A^{1s}$
ħω
ħω
PL
GS
c
$X_A^{1s}$
$X_A^{2p}$
band gap
$E_{gap}$
detection
$X_A^{1s}$ Emision Intensity (arb. u.)
Linear. Polar. (%)
Two-Photon Excitation Energy (eV)
d
SHG
2P-PL
e

**Figure 3**

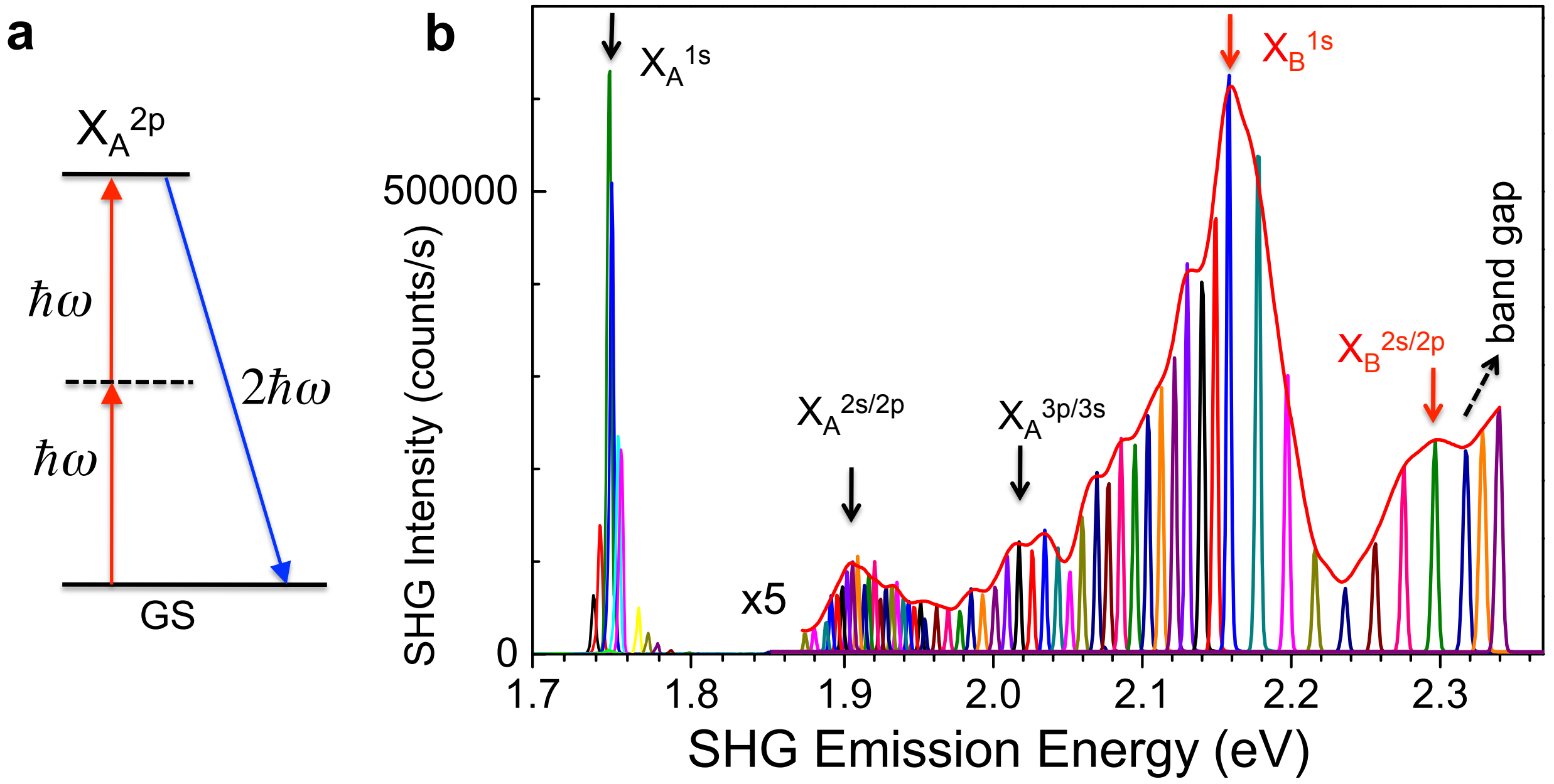

Figure 4

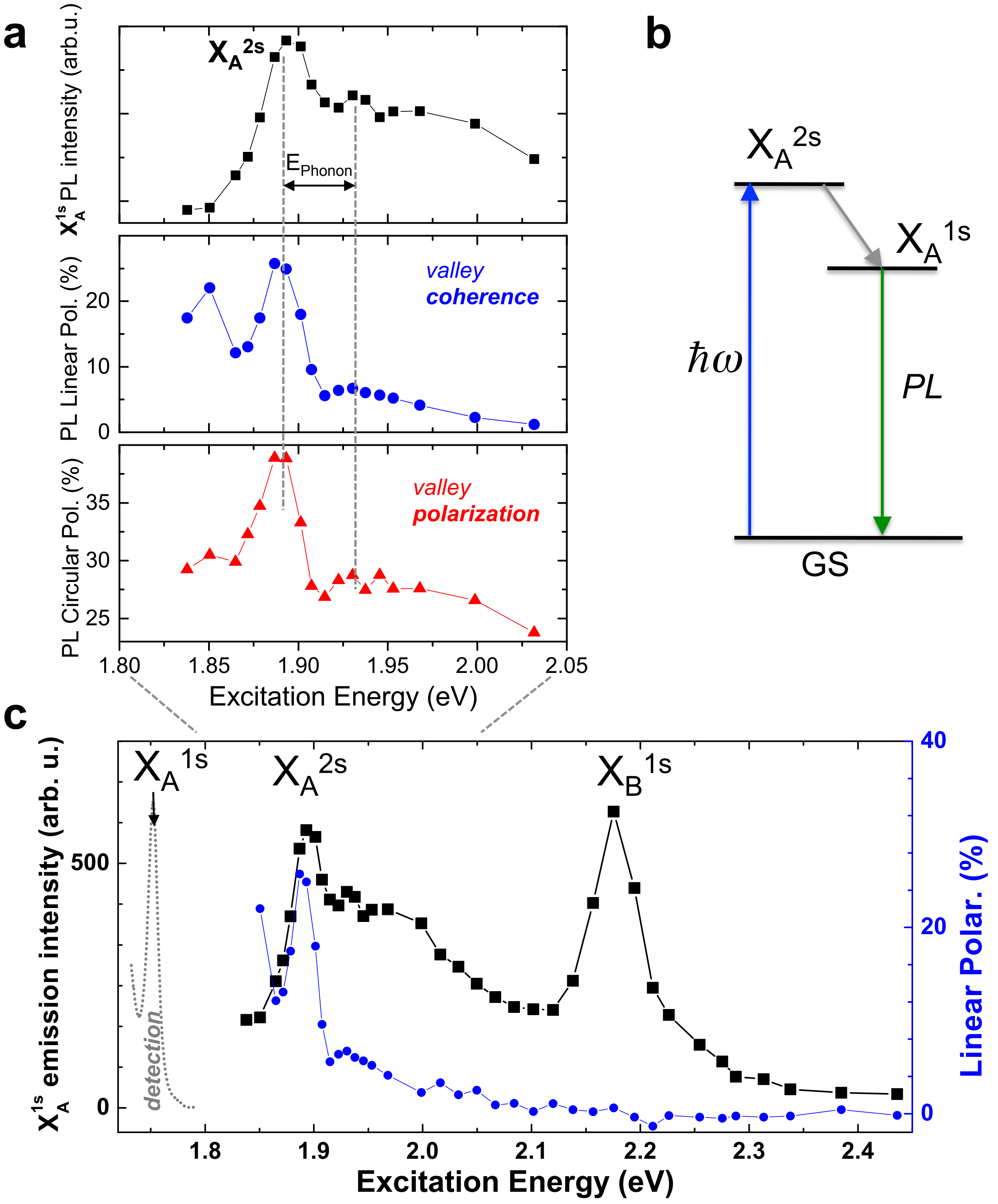

**Figure 5**

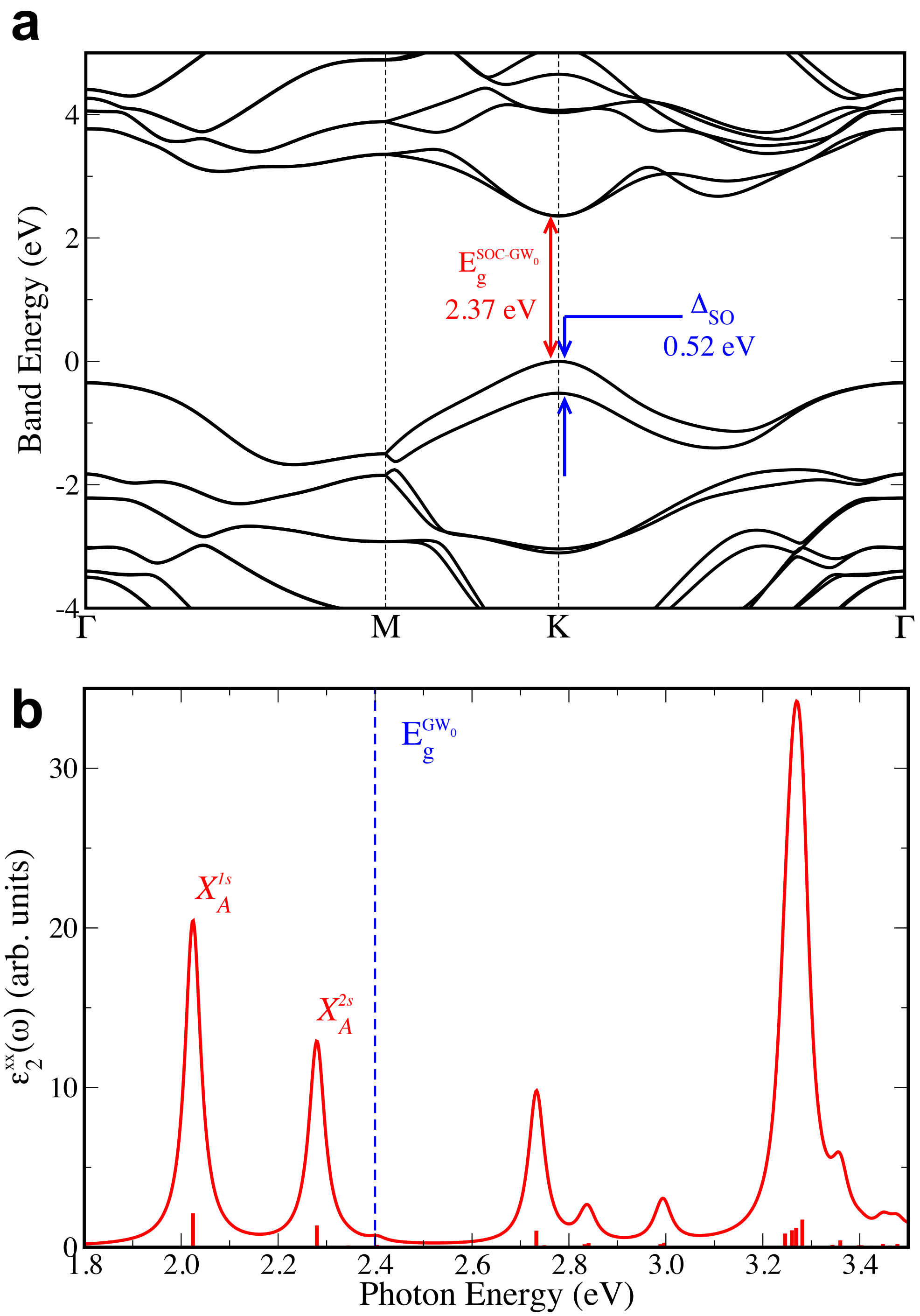

**Non-linear Optical Spectroscopy of Excited Exciton States**
**for Efficient Valley Coherence Generation in $WSe_2$ Monolayers**

***Supplementary Information***

G. Wang X. Marie, I. Gerber, T. Amand, D. Lagarde, L. Bouet, M. Vidal, A. Balocchi & B. Urbaszek

*Université de Toulouse, INSA-CNRS-UPS, LPCNO,*

*135 Av. de Rangueil, 31077 Toulouse, France*

**Exciton-phonon coupling and Raman resonances in the photoluminescence spectra of mononlayer $WSe_2$**

Figure S1. We display the results of three different runs of experiments corresponding to figure 2a. In addition to the main $X_A^{2p}$ resonance the observation of phonon-assisted two-photon absorption s is reproducible in both intensity (left axis, black symbols) and linear polarization (right axis, blue symbols).

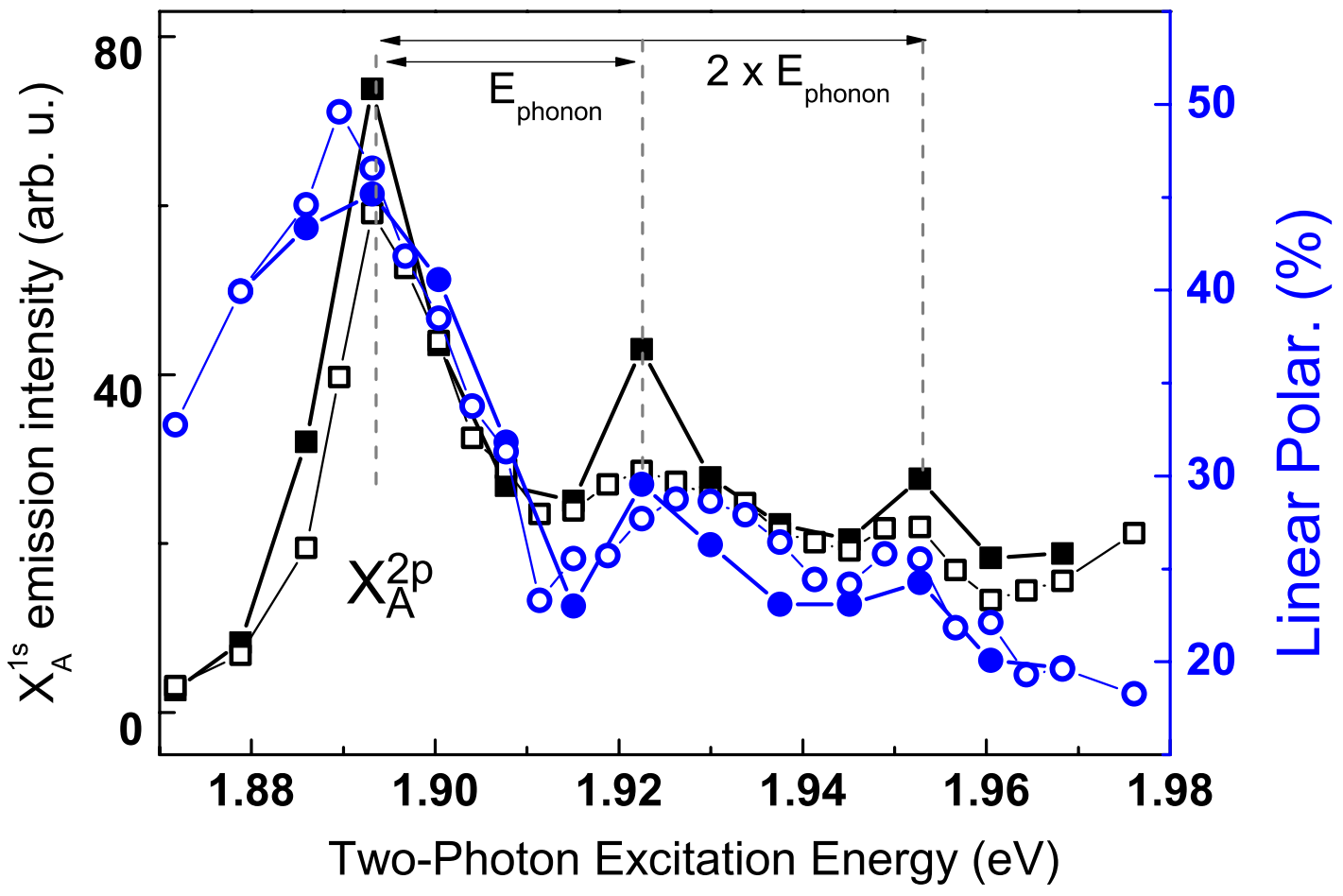

Figure S2. We show the photoluminescence (PL) of the neutral exciton state $X_A$ using the OPO for excitation. Superimposed on the PL signal we observe additional peaks at an energy of about 32meV below the excitation laser energy. These features are attributed to resonant Raman scattering, their spectral width is given by the pulsed excitation laser (~3meV).

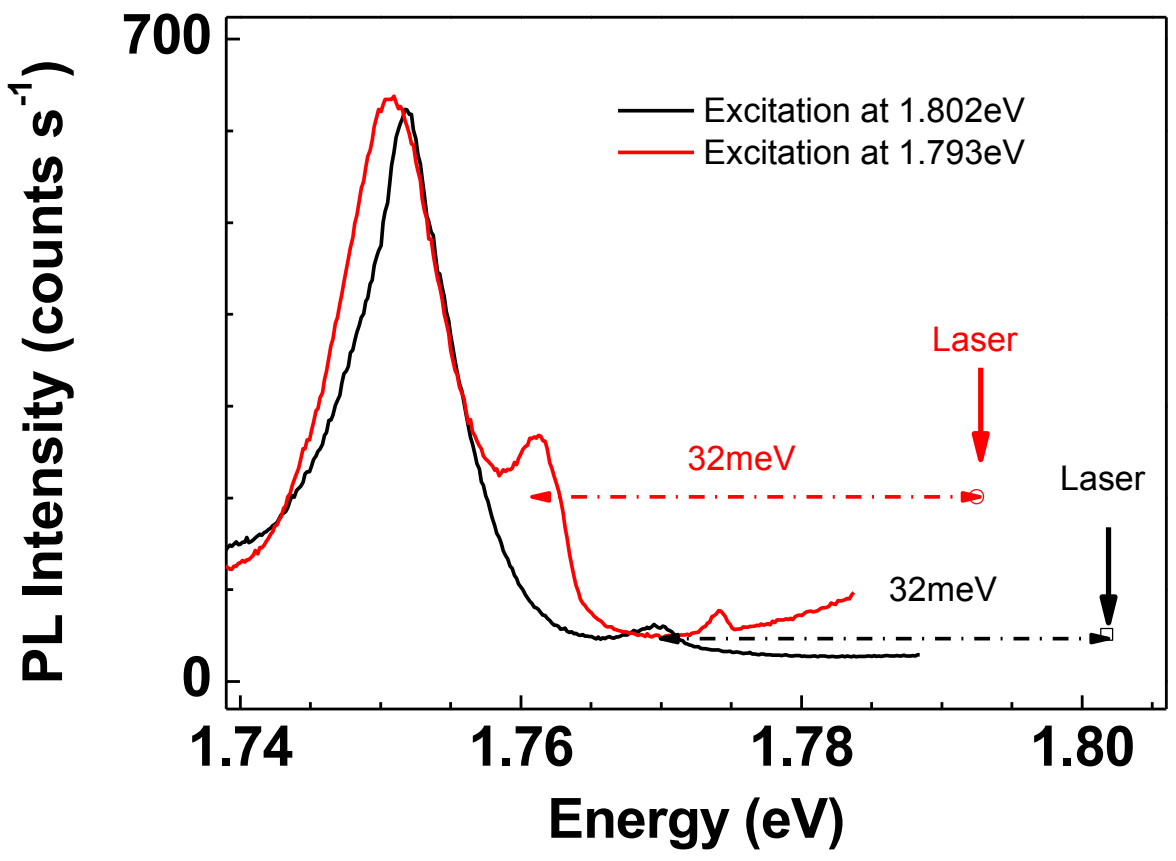

Figure S3. Raman signal on the same sample obtained with HeNe laser excitation. The spectrally narrow emission allows separating the two Raman modes $E^1_{2g}$ and $A_{1g}$.

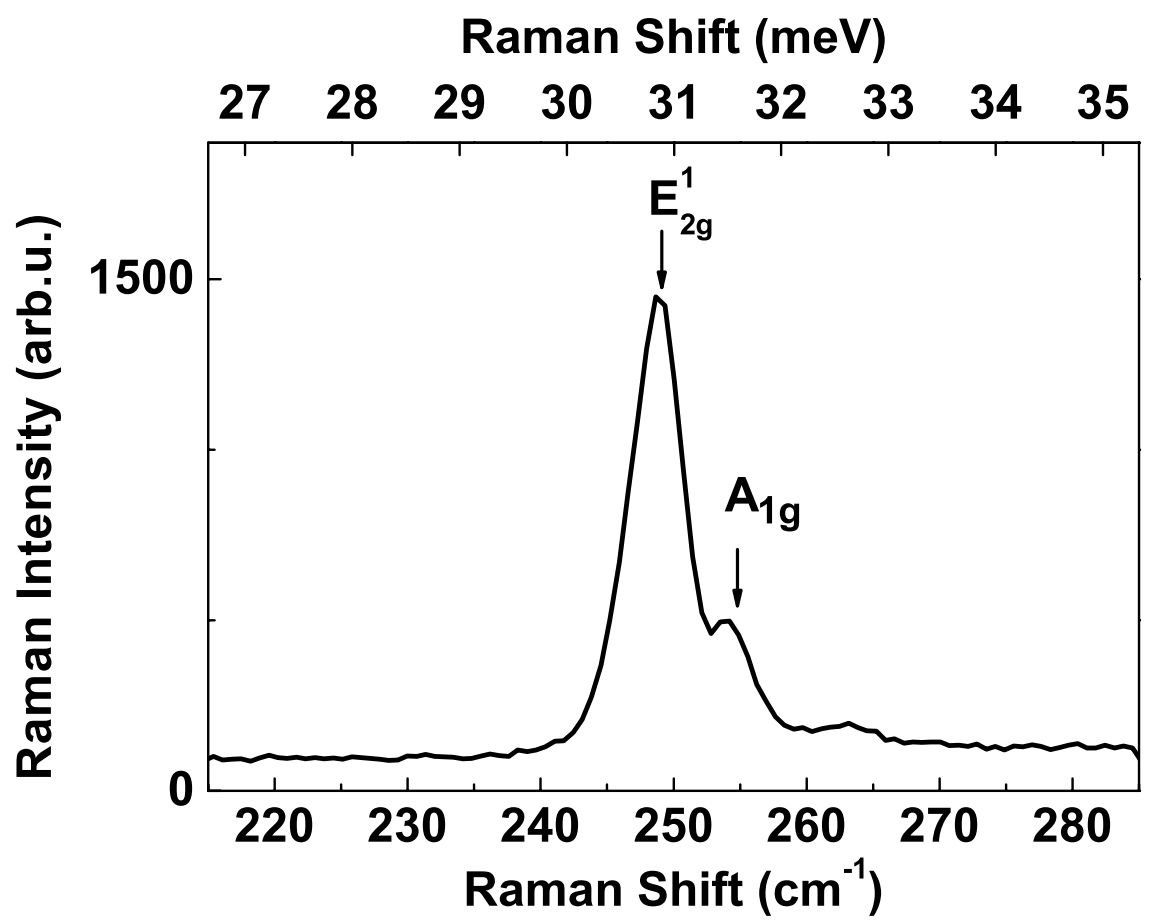

Figure S4. Following two-photon excitation, we observe for energies below 1.7eV unpolarized emission from localised states. The neutral exciton and the trion emission are marked as well as the SHG signal. This spectrum is an extended view of figure 1d. The exact shape and intensity of the localised emission differs between one and two-photon excitation.

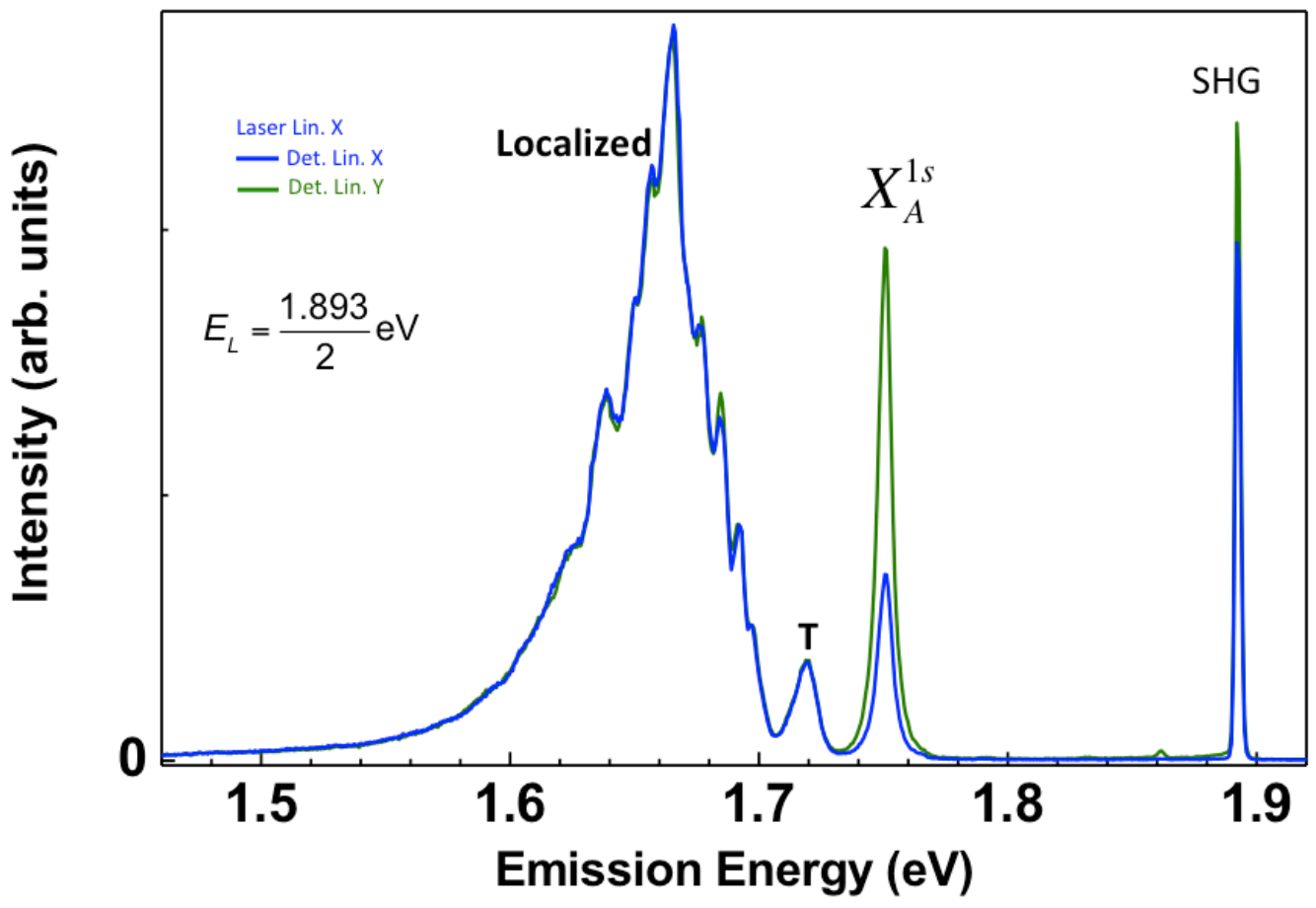

Below we first give a simple description of exciton states and their symmetry. We evaluate then the two photon absorption processes (TPA) in the interaction picture. We finally describe the second harmonic generation processes (SHG), first in a classical macroscopic approach in which we establish the selection rules. We then briefly describe the processes based on third order perturbation theory in the interaction representation. It confirms the selection rules, reveals the underlying mechanisms and allows us to qualitatively explain the resonances observed in the SHG spectroscopy.

**I. Exciton states and their symmetry.**

The exciton states description can be approximated in the envelope function approximation by the modified two-dimensional hydrogenic model. For the relative motion between electron and hole, we obtain for the radial part [1]:

$$R_{n,|l|}^{2D}(r) = N_{n,|l|} e^{-\rho_n} (2\rho_n)^{|l|} L_{n+|l|-1}^{2|l|}(2\rho_n) \qquad n \geq 1 \qquad |l| \leq n-1 \tag{1}$$

where $\rho_n \equiv \dfrac{1}{n-1/2} \dfrac{r}{a_{B,n}^*}$, $\mathbf{r} = \mathbf{r}_e - \mathbf{r}_h$, $a_{B,n}^*$ is the effective Bohr radius of state with principal quantum number *n*, and $N_{n,|l|}$ is a normalisation constant depending on *n* and the orbital quantum number *l*. Note in particular $N_{n,0} = \dfrac{\sqrt{2}}{(n-1/2)^{3/2}} \dfrac{1}{a_{B,n}^*}$. The 2D exciton relative motion wave function is thus $\psi_{n,l}^{2D}(r) = \dfrac{1}{\sqrt{2\pi}} R_{n,l}^{2D}(r) e^{il\varphi}$. The associated eigen-energies can be taken as $E_{n,|l|}^{2D} \approx -\dfrac{R_0}{\varepsilon^2(E_{n,0}^{2D})(n-1/2)^2}$, where $R_0 \equiv \dfrac{\mu e^4}{2\hbar^2}$ and $1/\mu = 1/m_c + 1/m_h$. Note the relation: $E_{n,0}^{2D} a_{B,n}^* = \dfrac{e^2}{2\varepsilon(E_{n,0}^{2D})}$. The phenomenological relative dielectric function $\varepsilon^2(\hbar\omega_{n,0}^{2D})$ has been introduced in order to take into account the observed departure from the usual two-dimensional hydrogenic series; it is assumed to have a slowly varying frequency dependence, so that it can be approximated by its value at the *nS* states for a given shell *n*. From the previous expression, we get:

$$\frac{E_{n,|l|}^{2D}}{E_{1,0}^{2D}} \approx -\frac{\varepsilon^2(E_{1,0}^{2D})}{\varepsilon^2(E_{n,0}^{2D})} \frac{1}{(2n-1)^2} \tag{2}$$

The exciton oscillator strength $f_{n,|l|}$ is proportional to $\left|\psi_{n,l}^{2D}(0)\right|^2$, so that $f_{n,|l|} = 0$ for $|l| > 0$, and for a given exciton bright state *nS* we have :

$$\frac{f_{n,0}}{f_{1,0}} \approx -\frac{\varepsilon^2(E_{1,0}^{2D})}{\varepsilon^2(E_{n,0}^{2D})} \frac{1}{(2n-1)^3} \tag{3}$$

The full exciton wave function is now approximated by:

$$\Psi^{\alpha,\beta}_{\mathbf{k}_G,n,l}(\mathbf{R}_G,\mathbf{r}) = \frac{1}{\sqrt{S}} e^{i\mathbf{k}_G.\mathbf{R}_G} \psi^{2D}_{n,l}(r)\zeta_c(z_e)\zeta_{A(B)}(z_h)u_{K_\alpha,c}(\mathbf{r})u^*_{K_\beta,\mathrm{v}}(\mathbf{r}) \quad (4)$$

where the functions $u_{K_\alpha,c(\mathrm{v})}(\mathbf{r})$ are the electron conduction (valence) Bloch functions close to the $K_\alpha$ ($\alpha$ = +,–) point of the Brillouin zone edge. The $\zeta_c(z_e)$ and $\zeta_{A(B)}(z_h)$ functions characterize the vertical confinement of the lowest conduction band ($c$) and of the $A$ or $B$ holes bands. These functions, which have a strongly pronounced maximum at the layer symmetry plane, have $\Gamma_1$ symmetry. The symmetry of the exciton states is now obtained as: $\Gamma_{exc} = \Gamma_{env} \otimes \Gamma_c \otimes \Gamma_h = \Gamma_{env} \otimes \Gamma_c \otimes \Gamma^*_{\mathrm{v}}$. We obtain thus for the $A$($B$) exciton states [2,3,4]:

| *type* | ($n$,\|$l$\|) | ($n$,\|$l$\|,$m$) $\Gamma_n$ | $K_+$ | $K_-$ |
|---|---|---|---|---|
| A | (1,0) | (1, 0, 0) $\Gamma_1$ | $\Gamma_2$, $\Gamma_4$ | $\Gamma_3$, $\Gamma_4$ |
| | (2,0) | (2, 0, 0) $\Gamma_1$ | $\Gamma_2$, $\Gamma_4$ | $\Gamma_3$, $\Gamma_4$ |
| | (2,1) | (2, 1,+1) $\Gamma_2$ | $\Gamma_3$, $\Gamma_5$ | $\Gamma_1$, $\Gamma_5$ |
| | | (2, 1,−1) $\Gamma_3$ | $\Gamma_1$, $\Gamma_6$ | $\Gamma_2$, $\Gamma_6$ |
| | | (2, 1, 0) $\Gamma_4$ | $\Gamma_1$, $\Gamma_5$ | $\Gamma_1$, $\Gamma_6$ |

| *type* | ($n$,\|$l$\|) | ($n$,\|$l$\|,$m$) $\Gamma_n$ | $K_+$ | $K_-$ |
|---|---|---|---|---|
| B | (1,0) | (1, 0, 0) $\Gamma_1$ | $\Gamma_2$, $\Gamma_6$ | $\Gamma_3$, $\Gamma_5$ |
| | (2,0) | (2, 0, 0) $\Gamma_1$ | $\Gamma_2$, $\Gamma_6$ | $\Gamma_3$, $\Gamma_5$ |
| | (2,1) | (2, 1,+1) $\Gamma_2$ | $\Gamma_3$, $\Gamma_4$ | $\Gamma_1$, $\Gamma_6$ |
| | | (2, 1,−1) $\Gamma_3$ | $\Gamma_1$, $\Gamma_5$ | $\Gamma_2$, $\Gamma_4$ |
| | | (2, 1, 0) $\Gamma_4$ | $\Gamma_3$, $\Gamma_5$ | $\Gamma_2$, $\Gamma_6$ |

**Table 1**: The A and B exciton (1s, 2s, and 2p) states and their representations in the $C_{3h}$ group. The Exciton valley $K_\alpha$ is by convention the valley of its constitutive conduction electron.

The exciton states with electron and hole in opposite valleys, which are not optically active, are not represented. We verified with DFT calculations that the optically active $\Gamma_4$ intra-valley modes have considerably weaker oscillator strength than the allowed modes $\Gamma_2$ in $K_+$ and $\Gamma_3$ in $K_-$. They are only dipole-allowed under $z$-polarised light ($z$ is perpendicular to the 2D layer plane). For A exciton, they are split from the latter mode by the spin-orbit interaction in the conduction bands ($\Gamma_7$ and $\Gamma_8$) by a few meV typically. Figure S5 shows for instance the one photon optical transitions for circularly polarised incident light with wave vector orthogonal to the $WSe_2$ flake, deduced from the above symmetry analysis.

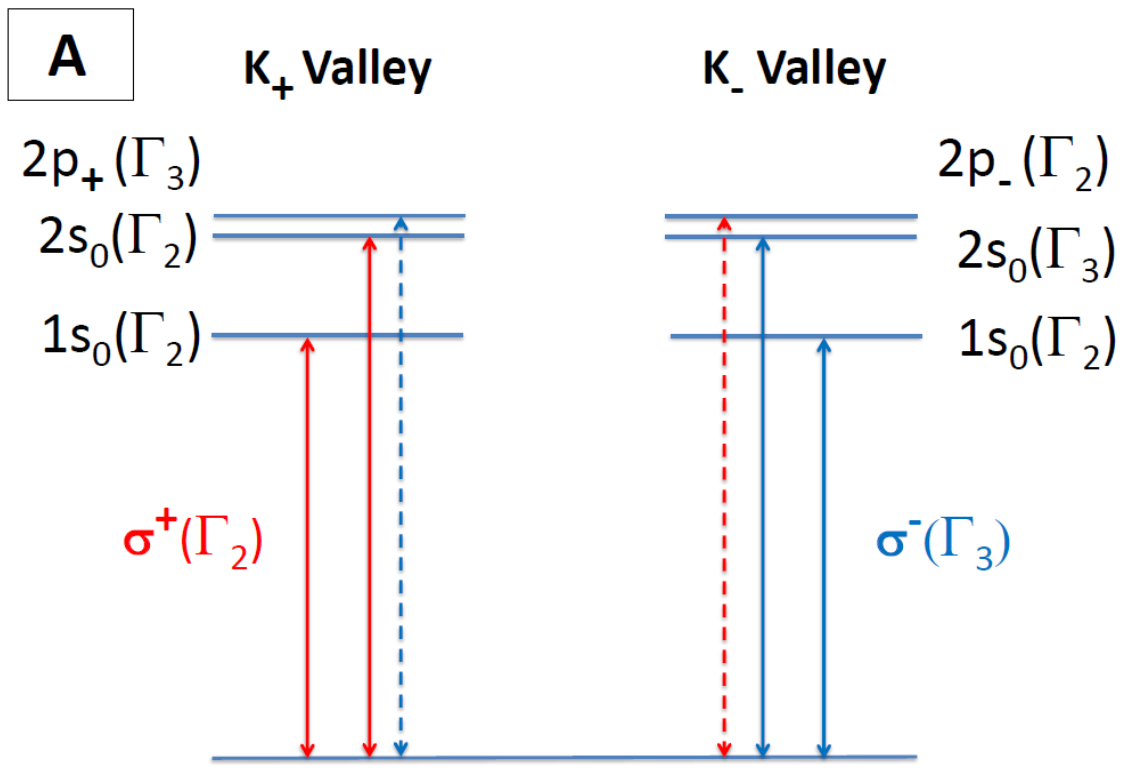

**Figure S5**: The allowed optical transitions for A exciton (1s, 2s, and 2p) under circularly polarized light propagating along the normal to the sample plane. The solid line arrows transitions results from dipolar coupling, while the dashed one are allowed according to magneto-dipolar transitions (see text, section III).

The magneto-dipolar transitions (see section III) have much smaller oscillator strength than the dipolar ones, the ratio between the two being of the order $a^*_{B,1}/\lambda_n$, $\lambda_n$ being the wavelength of the exciton transition. A similar scheme can be obtained for B-exciton.

## II. Two photon absorption process.

The two photon absorption, obtained following a perturbation scheme in the interaction representation in the dipolar approximation, is proportional to:

$$\frac{d\mathcal{P}_i}{dt}(2\omega)=\frac{2\pi}{\hbar}\left|\sum_{\alpha}\frac{\langle\Psi_{exc}|\hat{V}_i^{DE}|\Psi_\alpha\rangle\langle\Psi_\alpha|\hat{V}_i^{DE}|\varnothing\rangle}{E_\alpha-i\Gamma_\alpha-\hbar\omega}\right|^2\frac{\Gamma_{ex}/\pi}{(2\hbar\omega-E_{exc})^2+\Gamma_{ex}^2} \tag{5}$$

Here the index $\alpha$ runs over all virtual electron-hole pair states, and $\hat{V}_i^{DE}$ represents the exciton-photon interaction with photon mode of polarisation $\sigma^i$ in the dipolar approximation. In this approximation, the two-photon transitions from ground to 1*s*-states are forbidden under circularly polarized light. The transitions towards $2p_\pm$ states (*i.e.* (2, 1,±1) $\Gamma_{3(2)}$ exciton states) are allowed (see figure S6, left panel). The second step of TPA involves exciton dipolar transitions within *s*- and *p*-states of the electron-hole relative motion.

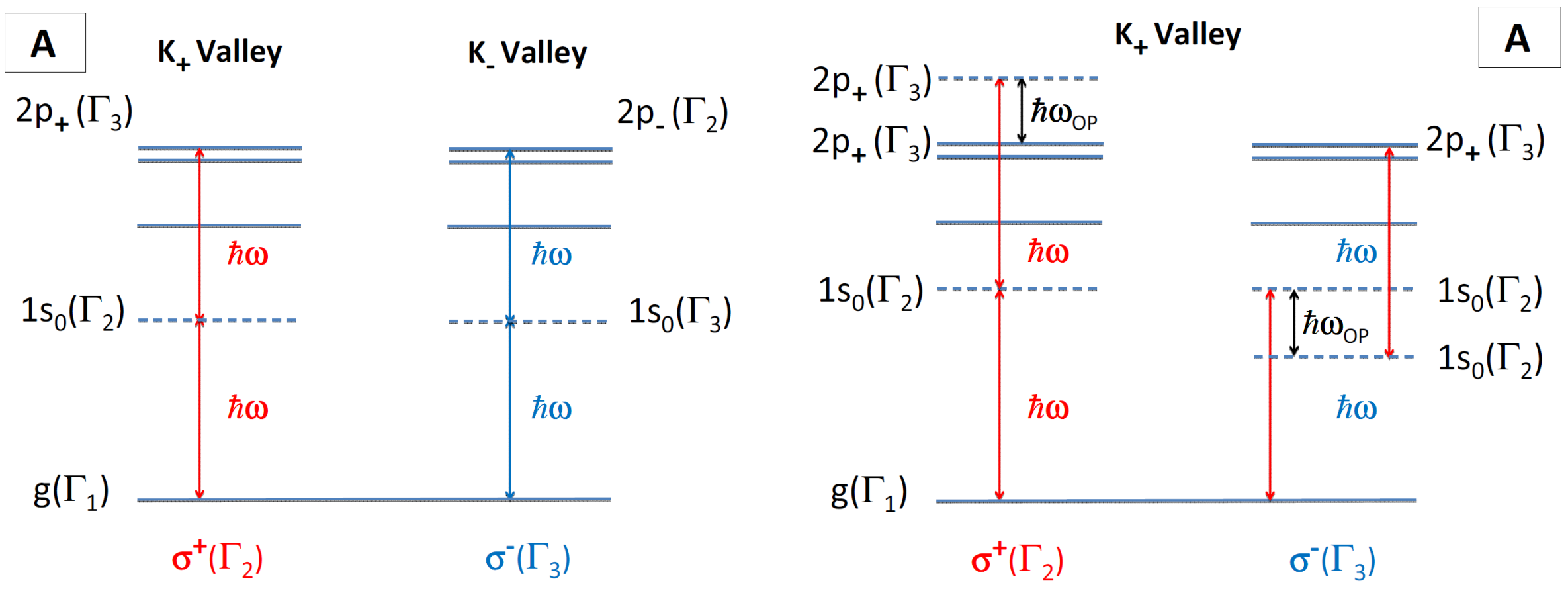

**Figure S6**: left: A-exciton absorption by resonant two-photon absorption process under circularly polarized light. Right: Optical phonon ($\hbar\omega_{OP}$) assisted two-photon absorption. The example shows the processes involving a $\Gamma_1$ ($A_{1g}$) optical phonon emission [5,6].

The 1-optical phonon assisted two-photon absorption (evidenced in figure 2a in the main text) can be obtained from a third order perturbation scheme. At low temperature it is proportional to (see figure S6, right panel):

$$\frac{d\mathcal{P}_i}{dt}(2\omega)=\frac{2\pi}{\hbar}\left|\sum_\alpha \frac{\langle\Psi_{exc}|\hat{V}^{em}_{ex-ph}|\Psi_\beta\rangle\langle\Psi_\beta|\hat{V}_i^{DE}|\Psi_\alpha\rangle\langle\Psi_\alpha|\hat{V}_i^{DE}|\varnothing\rangle}{(E_{ex}-E_\beta+i\Gamma_\beta-\hbar\omega)(E_\alpha-\hbar\omega)}+\frac{\langle\Psi_{exc}|\hat{V}_i^{DE}|\Psi_\beta\rangle\langle\Psi_\beta|\hat{V}^{em}_{ex-ph}|\Psi_\alpha\rangle\langle\Psi_\alpha|\hat{V}_i^{DE}|\varnothing\rangle}{(E_\beta-E_\alpha+i\Gamma_\alpha-\hbar\omega_{ph})(E_\alpha-\hbar\omega)}\right|^2\times$$
$$\times\frac{\Gamma_{ex}/\pi}{(2\hbar\omega-\hbar\omega_{ph}-E_{exc})^2+\Gamma_{ex}^2}\qquad(6)$$

Here, $\hat{V}^{em}_{ex-ph}$ represents the exciton-optical phonon interaction hamiltonian. The two-optical phonon assisted process can be obtained similarly from a fourth order perturbation scheme. The denominators of the expressions (5,6) explain the resonant character of the optical-phonon assisted absorption processes. The main contributions are given in this case by the dipolar matrix elements (and also the exciton-phonon matrix elements in the optical phonon assisted process) satisfying the selection rules.

**III. Second harmonic generation: macroscopic and microscopic approaches.**

Since $WSe_2$ monolayer crystal is non centro-symmetric, we expect significant second harmonic generation process to take place [7]. We describe first the phenomenological macroscopic equations relating the third-order optical susceptibility and the induced material polarisation at the origin of the emitted second harmonic electric field. Following ref. [7], the second order non-linear optical polarisation is given for a crystal of $D_{2h}$ symmetry by:

$$\begin{cases} P_x^{2\omega}=d_{11}(-2\omega;\omega,\omega)\left[\left(E_x^\omega\right)^2-\left(E_y^\omega\right)^2\right] \\ P_y^{2\omega}=-2d_{11}(-2\omega;\omega,\omega)E_x^\omega E_y^\omega \\ P_z^{2\omega}=0 \end{cases}\quad \text{or:}\quad \begin{cases} P_x^{2\omega}=-2d_{22}(-2\omega;\omega,\omega)E_x^\omega E_y^\omega \\ P_y^{2\omega}=-d_{22}(-2\omega;\omega,\omega)\left[\left(E_x^\omega\right)^2-\left(E_y^\omega\right)^2\right] \\ P_z^{2\omega}=0 \end{cases}\qquad(7)$$

depending whether the $x$-axis being orthogonal to or in a mirror plane containing the $z$-axis. Since the $WSe_2$ monolayer thickness is much smaller than the coherence length, the SHG is not influenced by phase-matching conditions [7,9]. For circularly polarised light, we obtain respectively for the two possible orientations:

$$\begin{cases} P_x^{2\omega} \pm iP_y^{2\omega} = d_{11}(-2\omega;\omega,\omega)\left(E_x^{\omega} \mp iE_y^{\omega}\right)^2 \\ P_z^{2\omega} = 0 \end{cases} \quad \text{or:} \quad \begin{cases} P_x^{2\omega} \pm iP_y^{2\omega} = -id_{22}(-2\omega;\omega,\omega)\left(E_x^{\omega} \mp iE_y^{\omega}\right)^2 \\ P_z^{2\omega} = 0 \end{cases} \quad (8)$$

Since the two situations corresponds to a rotation of the system of $\pi/2$ about $z$-axis, we have indeed: $d_{22}(-2\omega;\omega,\omega) = -id_{11}(-2\omega;\omega,\omega)$. We will use in the following the first orientation convention. Since the flake thickness is much smaller than the coherence length, the SHG is not influenced by phase-matching conditions.

The electric field radiated at frequency $2\omega$ is collinear with the non-linear second-order polarisation. The selection rules for second harmonic generation are thus transparent from expressions (8), *i.e.* two identical circularly polarized photons of energy $\hbar\omega$ propagating normal to the flake can be up-converted towards a photon of energy $2\hbar\omega$ with *opposite* helicity. They correspond to our experimental observations in the main text.

For linearly polarised incident light, the selection rules are more involved, and follow equation (7). The non-linear coefficient is in principle frequency dependent, since the considered crystal is dispersive, particularly close to exciton resonances. The following microscopic description of the process, together with the optical selection rules, gives a hint of the frequency dependence of the non-linear susceptibility. Taking now the polarisation axis ($\mathbf{x}'$,$\mathbf{y}'$) parallel or orthogonal to the laser polarisation direction $\mathbf{e}_\theta$, so that $\mathbf{E}^{\omega} \equiv \mathcal{E}_0\ \mathbf{e}_{\theta}$, we obtain:

$$\begin{cases} P_{x'}^{2\omega} = d_{11}(-2\omega;\omega,\omega)\left(\mathcal{E}_0^{\omega}\right)^2 \cos(3\theta) \\ P_{y'}^{2\omega} = -d_{11}(-2\omega;\omega,\omega)\left(\mathcal{E}_0^{\omega}\right)^2 \sin(3\theta) \\ P_z^{2\omega} = 0 \end{cases} \quad (9)$$

The intensity of the linear components along ($\mathbf{x}'$,$\mathbf{y}'$) axis are thus proportional to $I_{x'} \propto \cos^2(3\theta) = (1+\cos 6\theta)/2$ and $I_{y'} \propto \sin^2(3\theta) = (1-\cos 6\theta)/2$ respectively. The corresponding two polar diagrams have thus a $\pi/3$ angular periodicity, and are dephased by the angle $\pi/6$ (see figure S7). The observed 6-leaf flower diagram for $I_{x'}$ in the main text follows, as also recently shown [8,9].

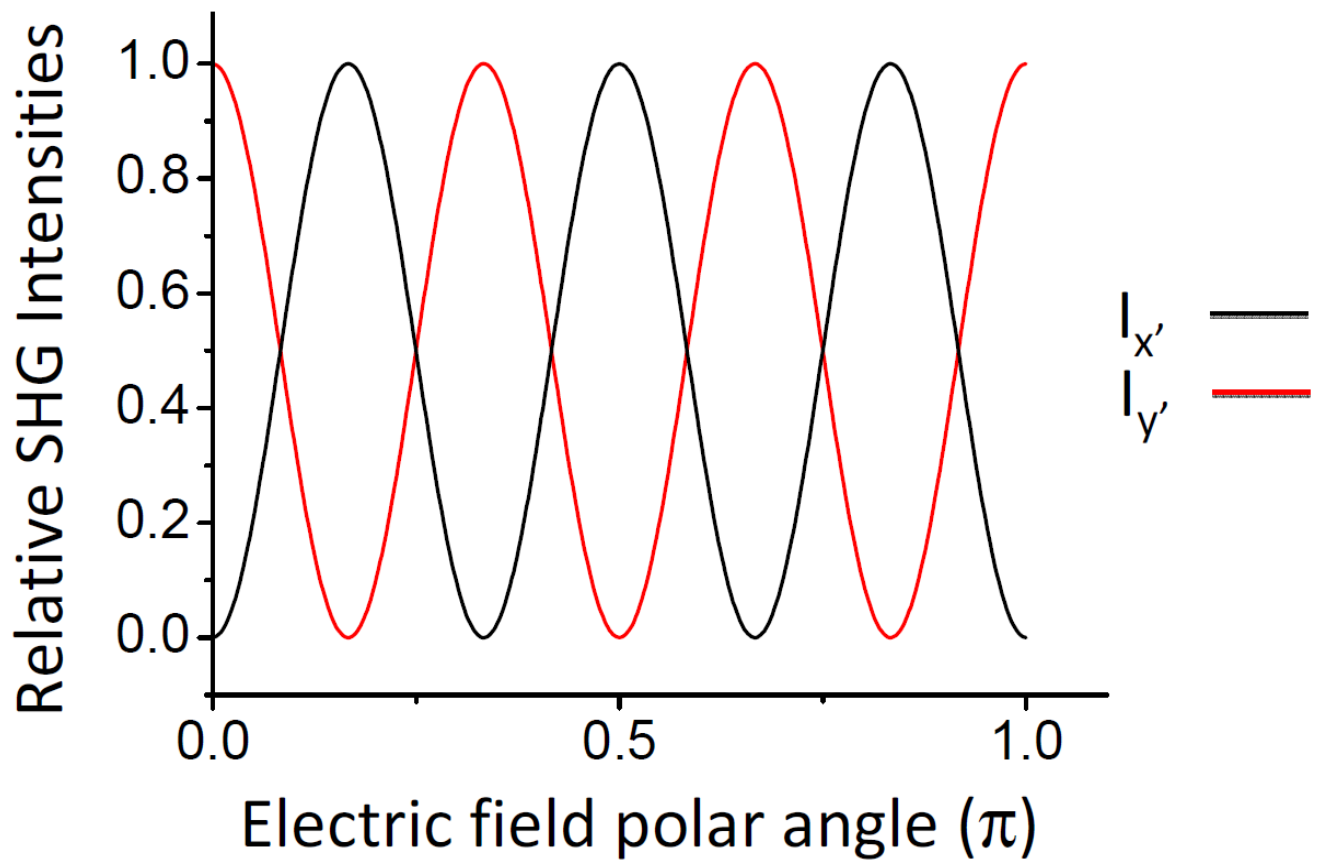

**Figure S7**: Intensity of the second harmonic signal co- or cross-polarized with a linearly polarized incident light which polarization vector makes an angle $\theta = (\mathbf{e}_x, \mathbf{e}_x)$ with respect to the x crystallographic axis.

We turn now to the microscopic approach. The non-linear dielectric susceptibility tensor $\chi^{(3)}$ can be evaluated on the ground of time dependent third-order perturbation theory in the interaction representation. Its components, close to resonance with a given $|\Psi_{exc}\rangle$ exciton state are proportional to [10]:

$$\chi^{(3)}_{ijk}(-2\omega;\omega,\omega) \propto \langle \varnothing | \hat{V}_i^{2\omega} \frac{|\Psi_{exc}\rangle\langle\Psi_{exc}|}{E_{exc} - 2\hbar\omega - i\Gamma_{exc}} \hat{V}_j^{\omega} \sum_{v} \frac{|\Psi_v\rangle\langle\Psi_v|}{E_v - \hbar\omega - i\Gamma_v} \hat{V}_k^{\omega} | \varnothing \rangle \tag{10}$$

where the term with $E_{exc} - 2\hbar\omega - i\Gamma_{exc}$ is dominant over similar terms obtained in the summation on all the other possible virtual exciton states. The tensor components $(i,j,k)$ may belong to the $\{x,y,z\}$ set, or to the $\{+,-,0\}$ standard set where +(–) correspond to right(left) circularly polarized components and $\{0\}$ to the $z$-polarized one. The imaginary $i\Gamma_v$ terms denote the decay of the virtual exciton states of energy $E_v$ due to scattering out the state. It is necessary here to develop the exciton-photon interaction hamiltonian up to first order in the light wage vector. For incident light normal to the sample plane, we have [11]:

$$\hat{V}_i^{\omega} = \hat{V}_i^{DE}(\omega) + \hat{V}_i^{QE}(\omega) + \hat{V}_i^{DM}(\omega) + h.c. \tag{11}$$

where the electric dipolar $\hat{V}_i^{DE}(\omega)$, electric quadrupolar $\hat{V}_i^{QE}(\omega)$ and magnetic dipolar interactions are respectively :

$$\hat{V}_{\perp}^{DE}(\omega) = -i\frac{\mathcal{E}_0}{2m_0c} e^{i\omega t} \mathbf{e}_{\perp}.\hat{\mathbf{p}}_{\perp} \tag{12}$$

$$\hat{V}_{\perp}^{QE}(\omega) = -\frac{q_z \mathcal{E}_0}{2m_0c} e^{i\omega t}\left[(\mathbf{e}_z.\hat{\mathbf{r}})(\mathbf{e}_{\perp}.\hat{\mathbf{p}}_{\perp}) + (\mathbf{e}_{\perp}.\hat{\mathbf{r}}_{\perp})(\mathbf{e}_z.\hat{\mathbf{p}}_z)\right] \tag{13}$$

$$\hat{V}_{\perp}^{DM}(\omega) = -\frac{q_z \mathcal{B}_0}{2m_0} e^{i\omega t} \mathbf{e}_z . \left[ \mathbf{e}_{\perp} \times (\hat{L} + 2\hat{S}) \right] \tag{14}$$

Here, $\mathbf{e}_{\perp}$ represents the light in-plane polarisation vector, $q_z = q = \omega / c$, $\mathcal{B}_0 = \mathcal{E}_0 / c$. Since the electric quadrupolar interaction is odd with respect to *z*, and the exciton wave function is even, the corresponding matrix element $\left\langle \Psi_{v'} \middle| \hat{V}_{\perp}^{QE} \middle| \Psi_{v} \right\rangle$ vanishes. As for the magnetic dipolar interaction, only the orbital part of the Hamiltonian is efficient here, since the spin-dependent part couples to non-optically active exciton states. The retained optical interactions are represented on figure S5. The most efficient sequences in (10) are finally the one containing two electric and one magnetic dipolar factors, with one of the dipolar electric factor corresponding to a 1*s* to 2*p* exciton internal transition. For circularly polarised light, the expressions (12,14) take the form:

$$\hat{V}_{\perp}^{DE}(\omega) = -i \frac{\mathcal{E}_0}{4m_0 c} e^{i\omega t} \left[ (e_x - ie_y)(\hat{p}_x + i\hat{p}_y) + (e_x + ie_y)(\hat{p}_x - i\hat{p}_y) \right] \tag{15}$$

$$\hat{V}_{\perp}^{DM}(\omega) = i \frac{q_z \mathcal{B}_0}{2m_0} e^{i\omega t} \left[ (e_x - ie_y)\hat{L}_+ + (e_x + ie_y)\hat{L}_- \right] \tag{16}$$

with $\hat{L}_{\pm} = \hat{L}_x \pm i\hat{L}_y$, so that for pure $\sigma^{\pm}$ light, $\hat{V}_{\pm}^{DE}(\omega) = -i \frac{\mathcal{E}_0}{4m_0 c} e^{i\omega t} (\hat{p}_x \pm i\hat{p}_y)$, and $\hat{V}_{\pm}^{DM}(\omega) == i \frac{q_z \mathcal{B}_0}{2m_0} e^{i\omega t} \hat{L}_{\pm}$. The figure S8 below shows typical examples of possible processes where the radiated light is resonant with the 1*s* (left) or 2*p* exciton (right).

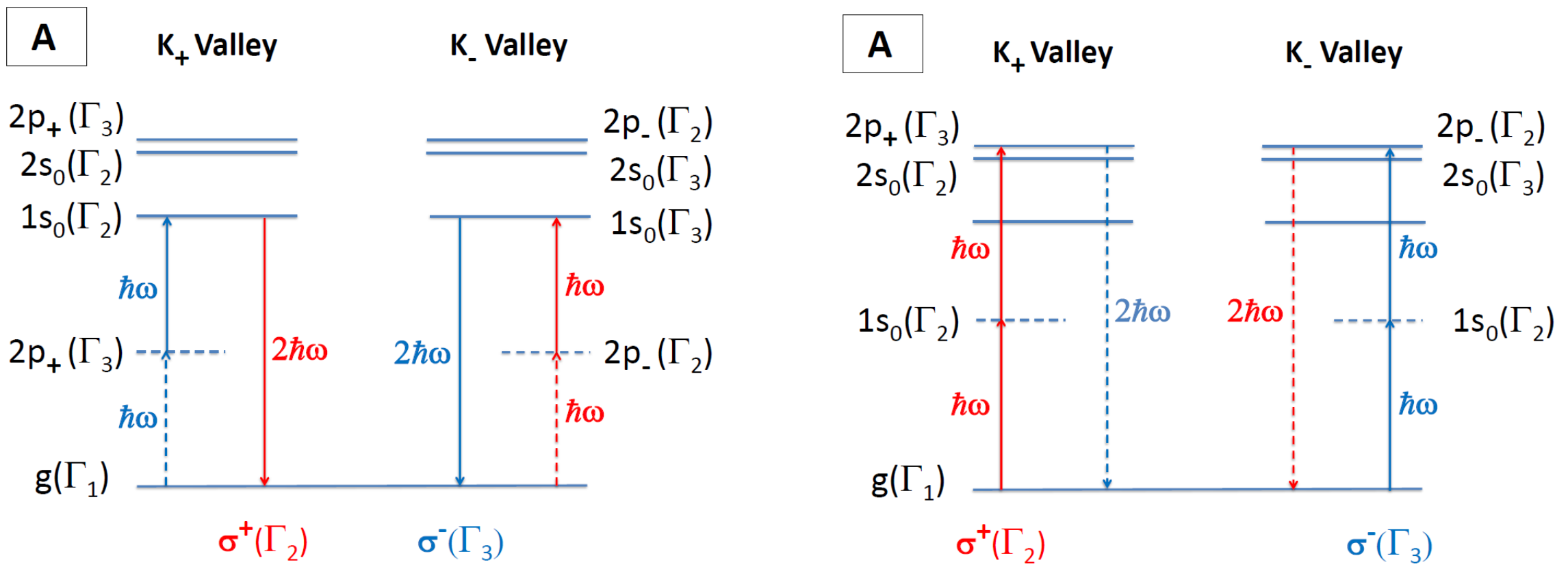

**Figure S8**: Second harmonic generation process with the $2\hbar\omega$ emitted photon in resonance with the A-1$s_0$ exciton states (left), or with the A-2$p_{\pm}$ exciton (right). The solid arrows represent electric dipolar interactions, while the dashed one the magnetic dipolar transitions. Note that the selection rules deduced from the macroscopic description are indeed obtained again.

For instance, on the left panel in the $K_+$ valley, the displayed sequence corresponds first to the excitation from the ground state of a virtual 2$p_+(\Gamma_3)$ state under the effect of magnetic dipolar interaction with one σ- photon. Second, an electric dipolar internal interaction couples this $2p_+(\Gamma_3)$ virtual state to the virtual $1s\,(\Gamma_2)$ with a second σ- photon. Finally, the dipolar interaction couples the $1s\,(\Gamma_2)$ virtual state back to the ground state $|\varnothing\rangle$. Retaining the most important terms, we obtain from (10,15,16) the approximate expression:

$$\chi^{(3)}_{+,-,-}(-2\omega;\omega,\omega) \propto \langle\varnothing|\hat{V}_+^{DE\dagger}(2\omega)|\Psi_{1s_0}\rangle \frac{\langle\Psi_{1s_0}|\hat{V}_-^{DE}(\omega)|\Psi_{2p_+}\rangle}{E_{1s_0}-2\hbar\omega-i\Gamma_{1s_0}} \frac{\langle\Psi_{2p_+}|\hat{V}_-^{DM}(\omega)|\varnothing\rangle}{E_{2p_+}-\hbar\omega-i\Gamma_{2p_+}}$$

*i.e.*

$$\chi^{(3)}_{+,-,-}\left(-\omega_{1s_0}^{\sigma^+};\frac{\omega_{1s_0}^{\sigma^-}}{2},\frac{\omega_{1s_0}^{\sigma^-}}{2}\right) \propto i\frac{\langle\varnothing|\hat{V}_+^{DE\dagger}(2\omega)|\Psi_{1s_0}\rangle\langle\Psi_{1s_0}|\hat{V}_-^{DE}(\omega)|\Psi_{2p_+}\rangle\langle\Psi_{2p_+}|\hat{V}_-^{DM}(\omega)|\varnothing\rangle}{\Gamma_{1s_0}(E_{2p_+}-\hbar\omega)} \quad (17)$$

since $\Gamma_{2p_+} << E_{2p_+}-\hbar\omega$. Similar expressions hold when the generated photon is resonant with the exciton $2s$ state. The SGH scheme for the $K_-$ valley with σ- polarised incident light can also be derived in the same way.

We take now the second harmonic process with the generated photon in resonance with the $2p$-exciton in $K_+$ valley seen on the right panel of fig. S8. This process corresponds first to the excitation from the ground state of a $1s\,(\Gamma_2)$ virtual state followed by the excitation to the $2p_+(\Gamma_3)$ virtual state under the effect of an electric dipolar internal interaction. Finally the magnetic dipolar interaction with one σ- photon drives the system back to the ground state. We obtain now:

$$\chi^{(3)}_{-,+,+}(-2\omega;\omega,\omega) \propto \langle\varnothing|\hat{V}_-^{DM\dagger}(2\omega)|\Psi_{1s_0}\rangle \frac{\langle\Psi_{2p_+}|\hat{V}_+^{DE}(\omega)|\Psi_{1s_0}\rangle}{E_{2p_+}-2\hbar\omega-i\Gamma_{2p_+}} \frac{\langle\Psi_{1s_0}|\hat{V}_+^{DE}(\omega)|\varnothing\rangle}{E_{1s_0}-\hbar\omega-i\Gamma_{1s_0}}$$

that is :

$$\chi^{(3)}_{-,+,+}(-\omega_{2p_+};\frac{\omega_{2p_+}}{2},\frac{\omega_{2p_+}}{2}) \propto i\frac{\langle\varnothing|\hat{V}_-^{DM\dagger}(2\omega)|\Psi_{2p_+}\rangle\langle\Psi_{2p_+}|\hat{V}_+^{DE}(\omega)|\Psi_{1s_0}\rangle\langle\Psi_{1s_0}|\hat{V}_+^{DE}(\omega)|\varnothing\rangle}{\Gamma_{2p_+}(E_{1s_0}-\hbar\omega)} \quad (18)$$

All these expressions confirm both the selection rules under circular polarisation deduced from the macroscopic approach, and bring in addition a qualitative interpretation for the peaks observed in the SHG spectroscopy shown in the main text whenever the generated photon is resonant with a real $1s$- or $2p$-exciton. The radiated intensities corresponding to expressions (13,14) are proportional to $\left|\chi^{(3)}_{\mp,\pm,\pm}(-2\omega;\omega,\omega)\right|^2$, which amplitudes are inversely proportional to $\Gamma_{2p/1s}$. Noticing that the involved matrix elements are of similar amplitude for resonant $1s$ and $2p$ processes, the fact that the exciton excited states can relax towards the $1s$ ones (see *e.g.* one-photon PLE experiments in main text), is consistent with the observation that the amplitude of the SHG resonant with 2s/$2p$ states is experimentally smaller than the one on the $1s$ states.

Finally, let us remark that SHG processes are also possible with the B excitons states according to similar schemes.

**Computational Details:**

In the paper, the atomic structures, the quasi-particle band structures and optical spectra are based on Density Functional Theory calculations performed using the VASP software [12]. The electron exchange and correlation are described by the PBE functional [13]. It uses the plane-augmented wave scheme [14] to treat core and valence electrons. Fourteen valence electrons are used to describe both the Mo and W atoms, while 6 electrons are included for the S and Se pseudo-potentials. A gaussian smearing of width equals to 0.05 eV is used to calculate partial occupancies. All atoms are allowed to relax with a force convergence criterion below 0.005 eV/Å. Spin-orbit coupling was also included non-self-consistently to determine eigenvalues and wave functions as input for the *GW* calculations. A tight electronic minimization tolerance of $10^{-8}$ eV is used here to determine with a good precision the large number of unoccupied states and the corresponding derivative of the orbitals with respect **k**. In this respect a total number of 256 bands is sufficient to converge the direct quasiparticle gap, within 0.05 eV, see table 2. Full-frequency-dependent *GW* calculations [15] are performed at the partially consistent $GW_0$ level; with 2 iterations of the *G* term leaving *W* at the DFT level, which is sufficient, since after 4 iterations, the changes in the quasiparticle band-gap in K is smaller than 10 meV. We have used the WANNIER90 code [16] and the VASP2WANNIER90 interface [17] to interpolate the band structures to a finer grid. The energy cutoff for the response function is set to be 270 eV. As explained in several recent studies [18,19], the distance between periodic images in *z* direction (*c* parameter) is a key parameter that influences the quasiparticle band-gap value. As shown in table 3, when keeping fixed the number of states (256) and the **k**-point mesh, the quasiparticle gap in Γ, vary significantly by 0.2 eV from 15.3 to 25.1 Å. Since the direct band-gap in K, is smaller (around 2.4 eV) one can expect that using a cell of 15.3 Å will underestimate the direct quasiparticle band-gap by roughly 0.1 eV.

Calculations of Bethe-Salpeter spectra are carried out on top of $GW_0$ wave functions, including the six highest valence bands and the eight lowest conduction bands. A complex shift of 0.02 eV is applied in all optical calculations, leading to a broadening of the theoretical absorption spectrum. For the Brillouin zone integration, a (11×11×1) Γ-centered **k**-point mesh is used for the calculations of quasiparticle band structures including spin orbit coupling (SOC), while a (21×21×1) grid is used to yield optical spectra, without spin orbit coupling. As explained in several recent studies [19-21], the **k**-point sampling is crucial in the determination of features in the imaginary part of the frequency dependent dielectric constant. It means that a compromise, due to computational feasibility between cell length, k-point sampling, number of states and incorporation of SOC has to be find. In this respect the figure S9 presents simulated spectra for $MoS_2$ monolayer, without SOC. Below the quasiparticle band-gap (2.6 and 2.7 on the (21x21x1) and (12x12x1) grid respectively), the effect of increase the vacuum length is to shift the peaks to smaller energies by 0.1 eV. At the same time if one increases the k-point sampling, a second peak appears, at 2.55 eV on the (12x12x1) mesh and at 2.5 eV on the finer grid. It provides a A-A' energy difference of 220 meV, in a reasonable agreement with the 320 meV found in Ref. [21]. Same behavior is obtained on $WSe_2$ case, as shown in Figure S10. Additionally we have tested influence of the vacuum length on spectra and clearly it is shown in Figure S11 that increasing the vacuum shifts to larger energy the peaks but leaves the differences between them untouched.

*Table 2: Calculated gaps of $WSe_2$ monolayer value at $\Gamma$ point for a calculation cell of length 15.3 Å, with a (11×11×1) grid as a function of the total number of states.*

| Number of Bands | 256 | 512 | 2048 |
|---|---|---|---|
| Gap [eV] | 4.35 | 4.32 | 4.32 |

*Table 3: Calculated gaps of $WSe_2$ monolayer value at $\Gamma$ point for a (11×11×1) grid and 256 states as a function of cell length in z direction.*

| Number of Bands | 15.3 | 19.0 | 25.1 |
|---|---|---|---|
| Gap [eV] | 4.35 | 4.44 | 4.55 |

*Figure S9: Frequency dependent imaginary part of the dielectric function $\varepsilon_2$ including excitonic effects, with different **k**-point meshes for $MoS_2$ ML.*

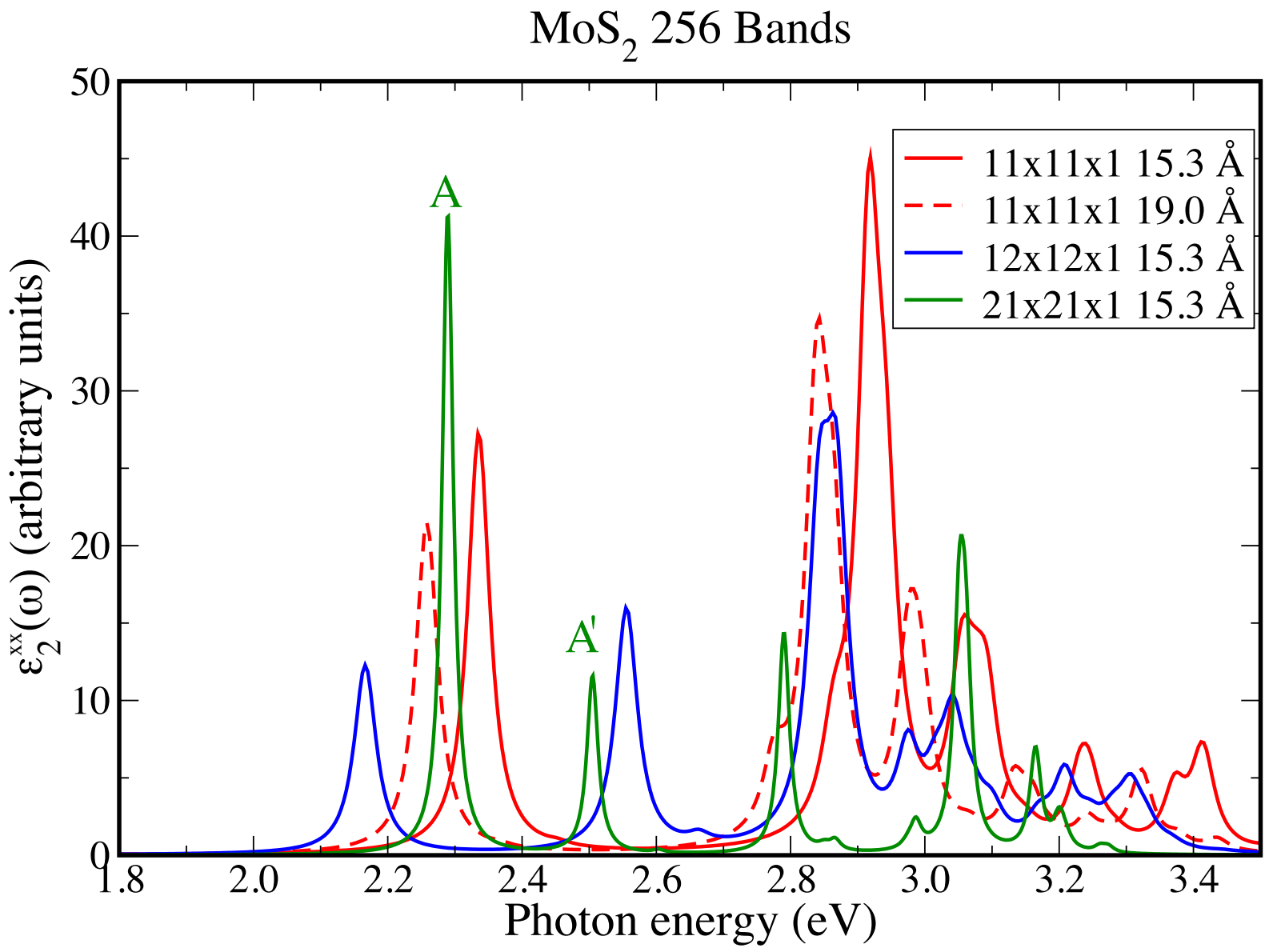

*Figure S10: Frequency dependent imaginary part of the dielectric function $\varepsilon_2$ including excitonic effects, with different **k**-point meshes for $WSe_2$ ML.*

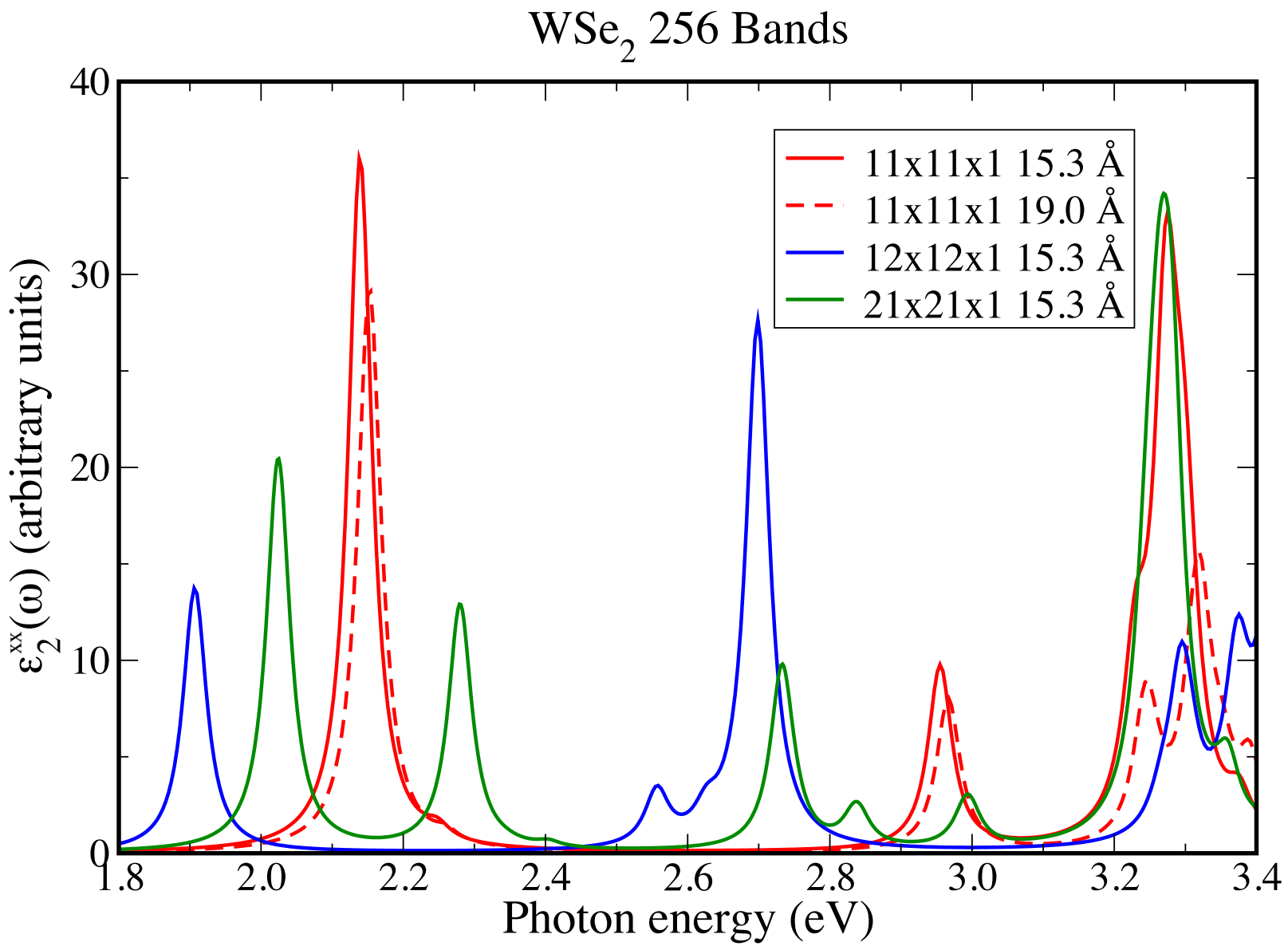

*Figure S11: Frequency dependent imaginary part of the dielectric function $\varepsilon_2$ including excitonic effects, with three different c parameters.*

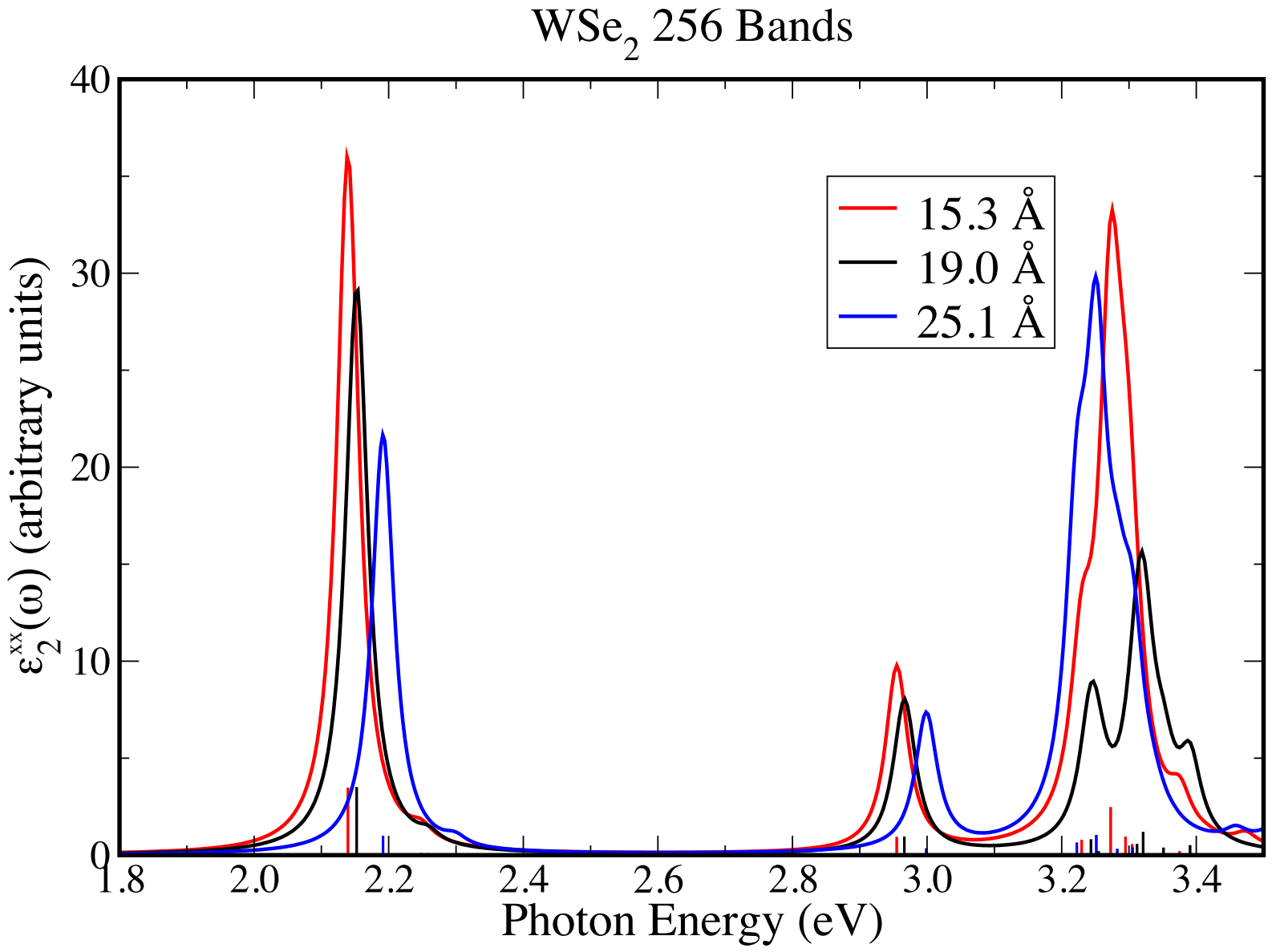

**Supplementary References:**